\pdfoutput=1
\documentclass[usenatbib]{mnras}
\usepackage[utf8]{inputenc}
\usepackage{graphicx}
\usepackage{amssymb}
\usepackage{mathtools}
\usepackage{amsmath}
\usepackage{color}
\usepackage{rotating}
\usepackage{pdflscape}
\usepackage{array}
\usepackage{multirow}
\usepackage{tabularx}
\usepackage{booktabs}
\usepackage{cite}
\usepackage{threeparttable}
\usepackage{caption}
\usepackage{subcaption} 
\title[Sub-mm Variability of AGB stars]{The Sub-mm Variability of IRC+10216 and $o$ Ceti}
\author[T. E. Dharmawardena et al.]{Thavisha E. Dharmawardena$^{1,2}$\thanks{tdharmawardena@asiaa.sinica.edu.tw}, Francisca Kemper$^{3,1}$, Jan G. A. Wouterloot$^{4}$, 
\newauthor
Peter Scicluna$^{1}$, Jonathan P. Marshall$^{1}$, Sofia H. J. Wallstr\"om$^{1,5}$\\
$^{1}$Academia Sinica Institute of Astronomy and Astrophysics, 11F of AS/NTU Astronomy-Mathematics Building, \\No.1, Sec. 4, Roosevelt Rd, Taipei 10617, Taiwan.\\
$^{2}$Graduate Institute of Astronomy, National Central University, 300 Zhongda Road, Zhongli 32001, Taoyuan, Taiwan.\\
$^{3}$European Southern Observatory, Karl-Schwarzschild-Str. 2, 85748 Garching b. M\"unchen, Germany \\
$^{4}$East Asian Observatory, 660 N A'ohoku Place, Hilo, Hawaii 96720, USA\\
$^{5}$Institute of Astronomy, KU Leuven, Celestijnenlaan 200D bus 2401, 3001 Leuven, Belgium \\
}

\begin{document}

\maketitle

\begin{abstract}

We present the sub-mm variability of two of the most well studied AGB stars, IRC+10216 and $o$ Ceti. The data are obtained at $450~\micron$ and $850~\micron$ as part of pointing calibration observations for the James Clerk Maxwell Telescope's SCUBA-2 instrument over a span of 7 years. The periods are derived using non-parametric methods, \texttt{Gatspy Supersmoother} and \texttt{P4J}  in order not to assume an underlying shape to the periodicity. These were compared to two Lomb-Scargle parametric methods. We find that for both sources and wavelengths the periods derived from all methods are consistent within $1\sigma$. The $850~\micron$ phase folded light curves of IRC+10216 show a time lag of $\sim 540$ days compared to its optical counterpart. We explore the origins of the sub-mm variability and the phase lag using radiative transfer models. Combining the modelling with findings in the literature, we find that the sub-mm emission and phase lag can be partially attributed to the dust formation/destruction cycle. A second, unknown mechanism must be invoked; we defer an investigation of the origin and nature of this mechanism to a future work.


\end{abstract}

\begin{keywords}
keywords: stars:AGB and post-AGB -- stars: variables: general -- stars: individual: IRC+10216 -- stars: individual: $o$ Ceti
\end{keywords}

\section{Introduction}
\label{sec:Introduction}

All Asymptotic Giant Branch (AGB) stars are variable with brightness variations occurring on timescales of a few hundred days. These long term periodic variations generally originate as a result of the stellar pulsations \citep{Hofner+Olofsson2018}. While there have been many studies into the variability of AGB stars, most of them are limited to the optical to mid-infrared. However, visual extinction at these wavelengths caused by large amounts of circumstellar dust around high mass-loss rate AGB stars makes their variability difficult to trace. 

Therefore, far-infrared (far-IR) and sub-millimetre (sub-mm) wavelength studies are important for determining the nature of the variability of AGB stars such as periods, properties and origins. Sub-mm wavelengths are almost entirely unaffected by extinction from circumstellar matter. Furthermore, changes in molecular bands that plague optical and near-infrared (near-IR) observations do not affect the sub-mm. Monitoring sub-mm changes in source brightness allows monitoring of dust formation and the dusty envelope, effectively studying the total star including its extended dusty circumstellar envelope.  

Located at a distance of $130\pm13$ pc \citep{McDonald2017} IRC+10216 (CW Leo) is one of the most extensively studied AGB stars. It is an archetypal C-rich pulsating variable star with a spherically symmetric envelope expanding at a nearly constant velocity \citep{Cernicharo2015}. Due to the heavy presence of dust in the stellar envelope it is optically thick at optical wavelengths and hence only visible as a dim point source. The thick dusty envelope also makes it one of the brightest near-IR sources in the sky. We have measured the envelope which extends out to $\sim 1\arcmin$ in the sub-mm continuum at $450\micron$ and $850\micron$ \citep{Dharmawardena2018}  using the James Clerk Maxwell Telescope's Sub-millimetre Common Use Bolometer Array-2 instrument \citep[SCUBA-2;][]{Holland2013}.

While IRC+10216 is well-studied in many aspects its variability has only been determined a handful of times in the optical and infrared using sparse, unevenly sampled data; sub-mm to radio wavelength determinations are even scarcer. \citet{Menten2006_CWLeoRadio} determined the variability of IRC+10216 in the radio continuum (1.3, 2.0 and 3.6 cm) using VLA observations over a span of three years. They estimated a period of $535 \pm 51$ days and argue that the emission arises from the stellar photosphere or a somewhat more extended radio-photosphere. \citet{Dehaes2007} determine a period of 655.3 days using 1.2 mm continuum observations. 

In a more recent study \citet{Groenewegen2012IRC+10216} determined a period of $639 \pm 4$ days from Herschel/SPIRE (250, 350 and 450 $\micron$) data. The authors also hypothesize that the variability of IRC+10216 is directly related to the central star. Further the authors observed the bow shock of IRC+10216 located at $\sim 3.5\arcmin$ from the central source. However this bow-shock was not detected in the deep sub-mm map and resulting surface-brightness profile presented in our previous work \citep{Dharmawardena2018}, as it is outside the region that SCUBA-2 is sensitive to. Therefore when we refer to the far-IR periodicity determined by \citet{Groenewegen2012IRC+10216}, we are only considering the period they determined for the central star and ignore any contribution from the bow shock.

\citet{Kim2015-CWLeoVariability} derived a period of $640 \pm 0.9$ days in the optical using data from \citet{Drake2014-CatlinaSurvey}. Similarly they derived periods in the JHKLM bands using light curve data from \citet{LeBertre1992-CstarLightCurves} and \citet{Shenavrin2011-StellarVariability}. Several reports which employed optical measurements show periods ranging from $630 - 650$ days \citep{Witteborn1980-CWLeoVariability, Dyck1991-CWLeoVariability, LeBertre1992-CstarLightCurves}. 

Additionally molecular line observations by \citet{Teyssier2015-CWLeoMolecularLines} using Herschel HIFI, SPIRE, and PACS show periods between $\sim 600-760$ days for the HCN in the circumstellar envelope of IRC+10216. Analysis of rotational lines emitted from several molecules and radicals from $\sim 8 - 116$ GHz using the IRAM 30 m telescope by \citet{Pardo2018_CWLeo_MolLines_1} also show variability in similar time scales. In addition, \citet{He2017_CWLeo_MolLines_2} and \citet{Fonfria2018_CWLeo_MolLines_3} show the variability in the line fluxes and in the relative line strengths of several rotational lines, including SiS and HCN.

The O-rich cool pulsating binary AGB star $o$ Ceti A, commonly known as $o$ Ceti or Mira is the archetypal Mira variable and is located at a distance of $91.7 \pm 10.2$ pc \citep{McDonald2017}. Mira variables are well-known Long Period Variables (LPVs) pulsating in the fundamental mode, having stable periods of several hundred days. This variability is attributed mainly to the typical stellar pulsations and variability of the stellar temperature \citep{Wong2016-MiraAtmosphere}. 

A well sampled unbroken optical light curve from 1902 to 2006 compiled by \citet{Templeton2009-MiraVariability} displays a stable long period of $333.09 \pm 0.04$ days. The low optical depth of $o$ Ceti facilitates the analysis of its light curves at optical wavelengths allowing the curves to be studied in detail for over a century. 

In this paper we present sub-mm variability studies of IRC+10216 and $o$ Ceti. The observations are obtained by the James Clerk Maxwell Telescope's (JCMT) Sub-millimetre Common-User Bolometer Array 2 (SCUBA-2) instrument at $450\micron$ and $850\micron$ \citep{Chapin2013, Holland2013}. Both stars are commonly used as SCUBA-2 pointing calibrators since its inception. 

In Sec. \ref{sec:Methods} of this paper we present the data and data reduction used. In this section we also describe the Point Spread Function (PSF) photometry techniques used to determine the stellar fluxes and the methods used to derive the period from the fluxes. The resultant light curves and period determination techniques are presented in Sec. \ref{sec:Results+Analysis}. In Sec. \ref{Sec:Discussion} and \ref{Sec:OrginOf_Submm} we discuss the periods derived, compare them to past publications and analyse the phase folded light curves to understand the possible origins of the sub-mm variability. In Sec. \ref{Sec:Conclusions} we present the conclusion of this paper.

The scripts and list of observations used required to reproduce the analysis, figures and tables presented in this paper is available from figshare at \url{https://figshare.com/projects/EvolvedStars_Submm_Variability/67040} under the project title \textit{EvolvedStars\_Submm\_Variability}. In addition, the tables containing the photometry data used to derive the light curves can be found in VizieR.


\begin{table*}
  \centering
  \caption{Summary of PSF Photometry Results}
    \begin{tabular}{llll}
    
    \hline
    \hline
    
    Source and Wavelength & Flux Range (Jy)  & Average Flux (Jy) & \multicolumn{1}{l}{$\sigma$ (Jy)} \\
    
    \hline
    
    IRC+10216 $450~\micron$ & $10 - 41$ & $25 \pm 0.20$ & 5 \\
    
    IRC+10216 $850~\micron$ & $\mathbf{8 - 14}$ & $11 \pm 0.04$ & 1 \\
          
    $o$ Ceti $450~\micron$ & $0.13 - 9.9$ & $2.7 \pm 0.04$ & 1.1 \\
    
    $o$ Ceti $850~\micron$ & $0.14 - 1.5$ & $0.6 \pm 0.01$ & 0.14 \\
    
        \hline
    \hline
    
    \end{tabular}%
  \label{Table:FluxSummary}%
\end{table*}%

\section{Methods}
\label{sec:Methods}

\subsection{Observations and Data Reduction}
\label{sec:Obs+Reduction}


IRC+10216 and $o$ Ceti are two sources frequently used to calibrate the pointing accuracy of the SCUBA-2 instrument. As a result there are $\sim 900$ and $\sim 700$ pointing observations for each source respectively from January 2011 -- December 2017. The observations add up to $\sim 45$ hours per source with individual observations ranging from $\sim 45$ seconds to $\sim 6$ minutes.  The full list of observations used along with relevant information such as observation date, time and length are provided in the GitHub repository. The observations were selected and reduced using the method presented in Sec. 2.2 and 2.3 by \citet{Dharmawardena2018}. The sources were reduced using Starlink version 2017A and the standard SCUBA-2 FCFs \citep{Dempsey2013}.

\subsection{PSF Photometry}
\label{sec:PSFphotMethod}


As described in \citet{Dharmawardena2018} the reduced observations on occasion had artifacts known as negative bowling and blooming where the background is over-reduced to zero around the central point source or was enhanced to be as bright as the central source giving rise to what appears to be extended emission. If we were to apply aperture photometry to determine fluxes, these artefacts would mean tailoring the aperture radii to each individual observation in order to avoid the affected regions. Since we have many hundreds of observations this is not feasible. 

To overcome this issues we opted to determine the source fluxes via PSF photometry. Both sources are shown to be extended in \citet{Dharmawardena2018}, and therefore we applied a modified PSF to include the flux of the extended circumstellar envelope. We convert the radial profile of the co-added maps from \citet{Dharmawardena2018} into a pseudo-PSF function and employ that to derive PSF photometry. The composite images removed the effects of negative bowling and blooming seen in the individual observations during co-addition process. By using the composite image radial profiles as our \textit{PSF} function we ensure the extended emission and the correct shape of the sources are taken into account. 

Having determined a suitable PSF function we use the PSF photometry routines in the \texttt{Astropy} affiliated package \texttt{photutils} \citep{Bradley2016-Photutils} to carry out PSF photometry on our observations. The program produces background-subtracted PSF photometry along with the uncertainties on the fluxes. We include a SCUBA-2 calibration uncertainty of $ 8\%$ in addition to the uncertainties returned by \texttt{photutils}.

\section{Results and Analysis}
\label{sec:Results+Analysis}

\subsection{Light Curves}

\begin{figure*}
\centering
\begin{subfigure}[b]{\textwidth}
  \centering
  \includegraphics[width=\textwidth]{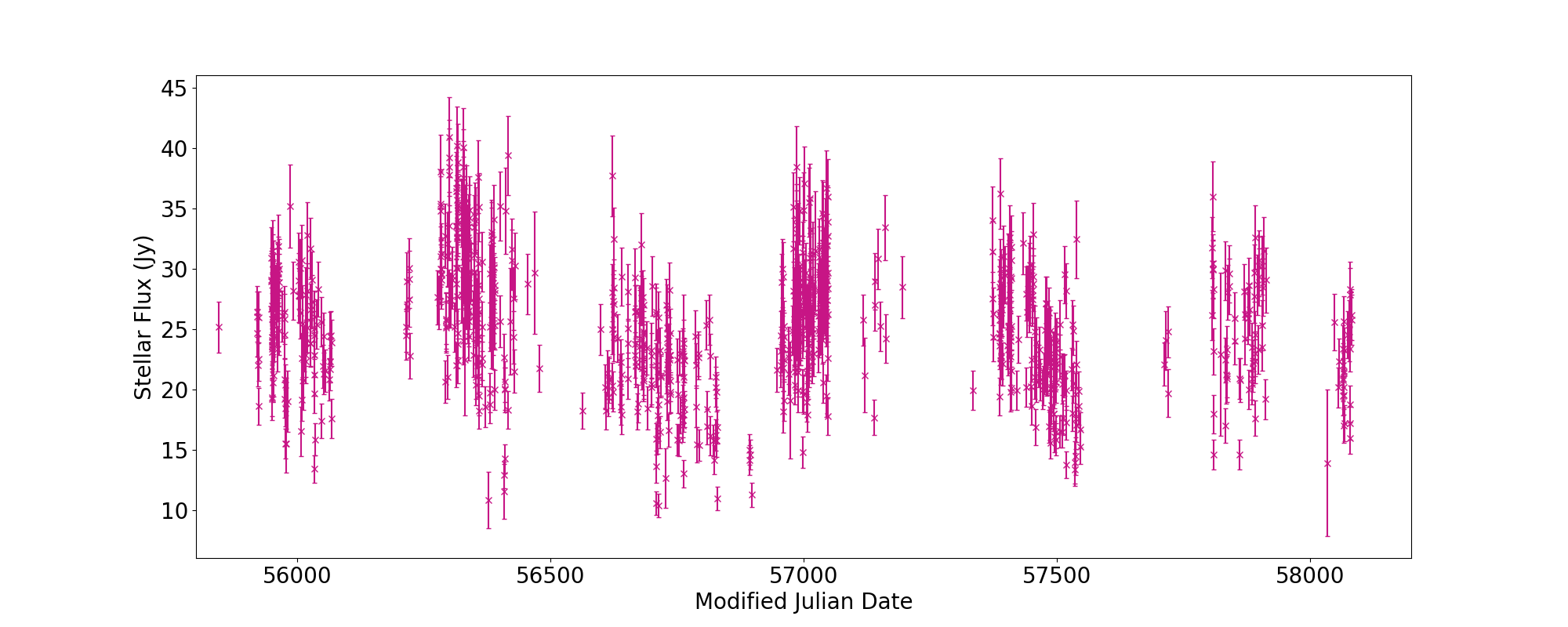}
  \subcaption{}
  \label{fig:CWLeo_450_LightCurve}
  \end{subfigure}
  
\begin{subfigure}[b]{\textwidth}
  \centering
  \includegraphics[width=\textwidth]{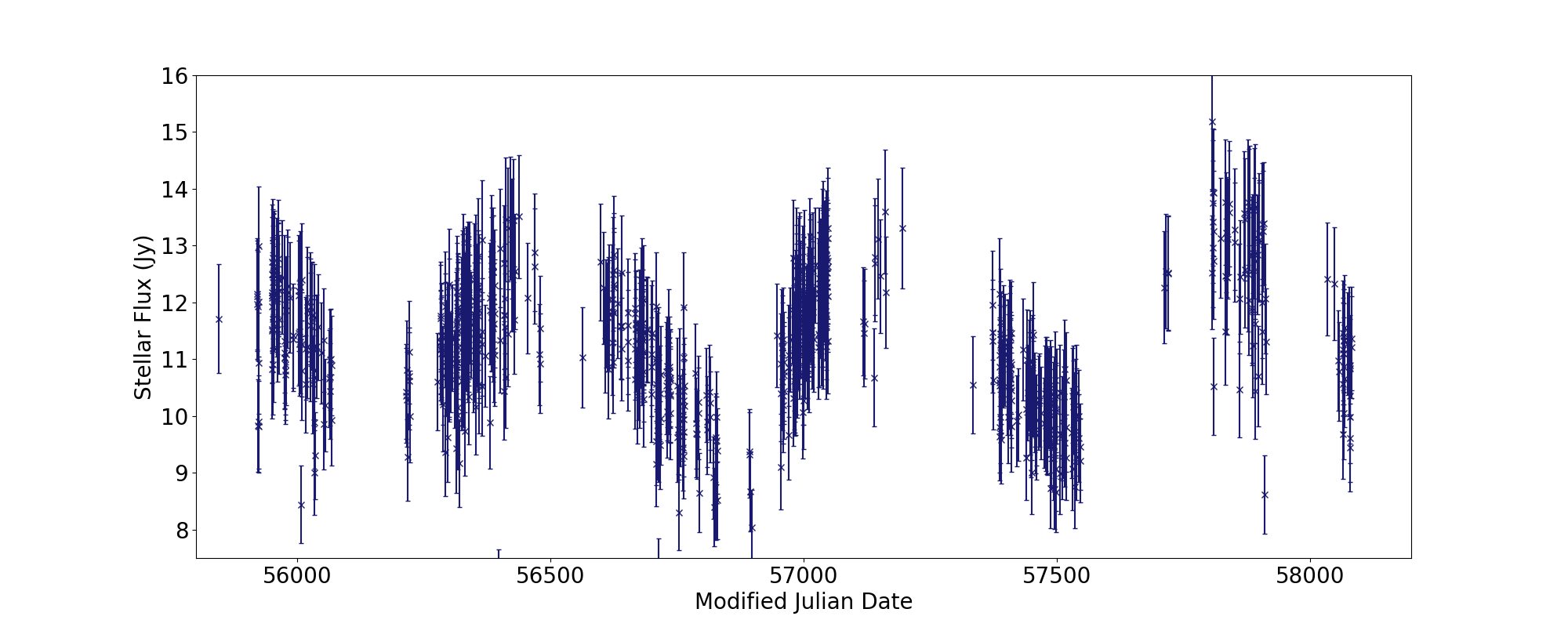}
  \subcaption{}
  \label{fig:CWLeo_850_LightCurve}
  \end{subfigure}

  \caption{Light Curves at $450~\micron$ (top) and $850~\micron$ (bottom) for IRC+10216. Samples of the photometry data required to produce these light curves are presented in Appendix~\ref{Appendix:LC_Data}. The full tables are given online as vizier tables.}
  \label{fig:CWLeo_AllLightCurves}
\end{figure*}

\begin{figure*}
\centering
\begin{subfigure}[b]{\textwidth}
  \centering
  \includegraphics[width=\textwidth]{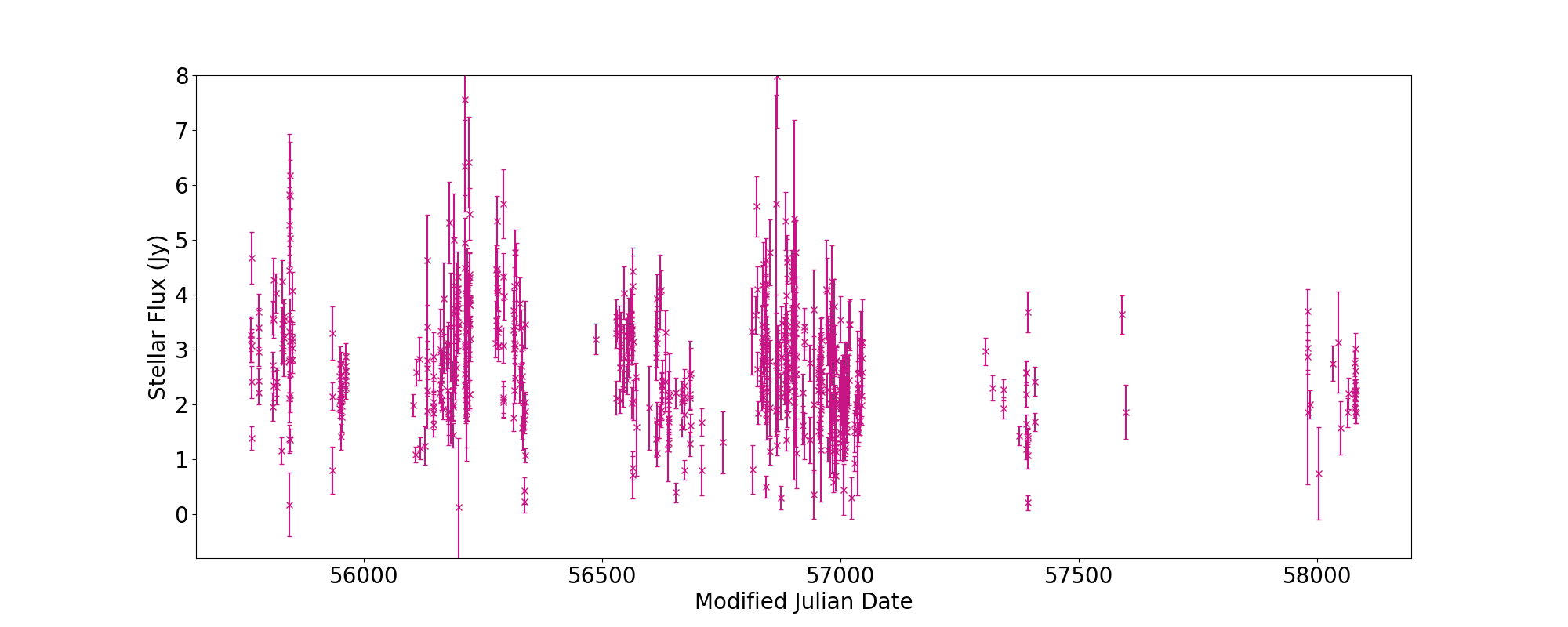}
  \subcaption{}
  \label{fig:$o$ Ceti_450_LightCurve}
  \end{subfigure}
  
\begin{subfigure}[b]{\textwidth}
  \centering
  \includegraphics[width=\textwidth]{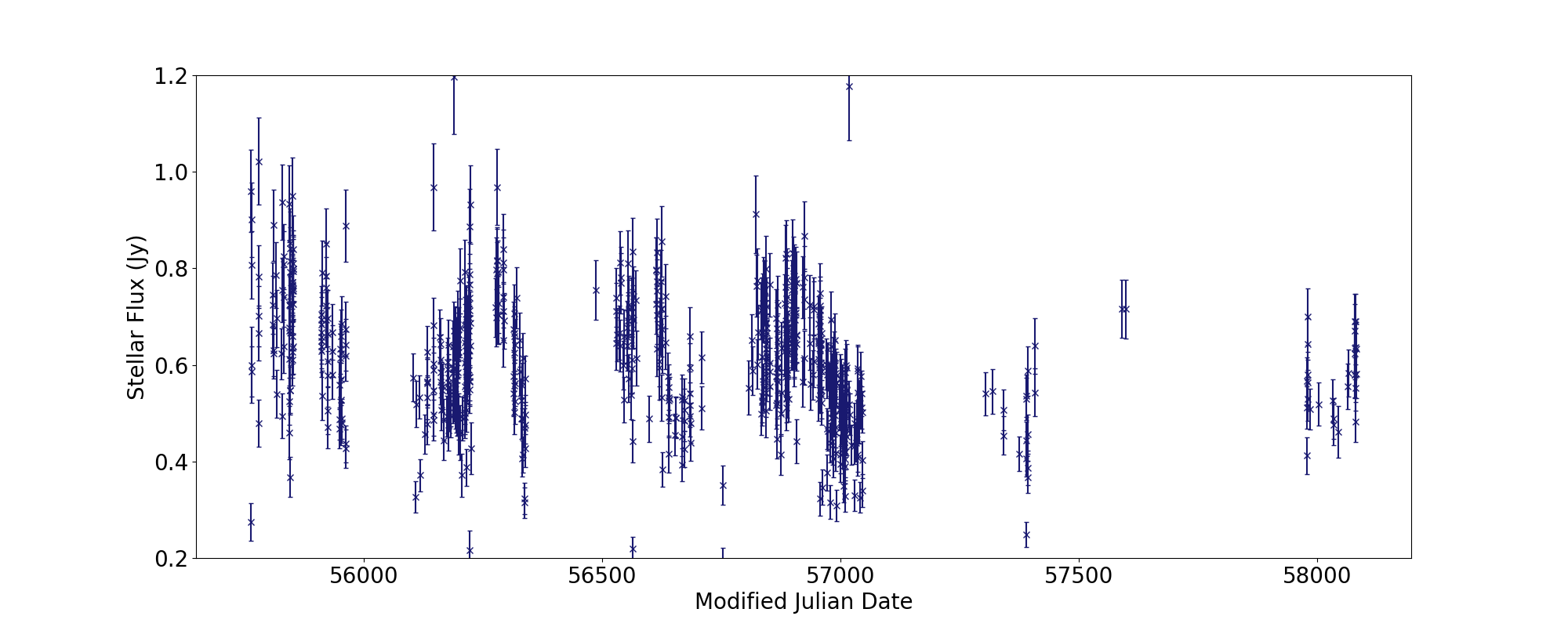}
  \subcaption{}
  \label{fig:$o$ Ceti_850_LightCurve}
  \end{subfigure}

  \caption{Light Curves at $450~\micron$ (top) and $850~\micron$ (bottom) for $o$ Ceti. Samples of the photometry data required to produce these light curves are presented in Appendix~\ref{Appendix:LC_Data}. The full tables are given online as vizier tables.}
  \label{fig:Mira_AllLightCurves}
\end{figure*}

Fig.~\ref{fig:CWLeo_AllLightCurves} presents the $450~\micron$ (top) and $850~\micron$ (bottom) light curves for IRC+10216, while Fig.~\ref{fig:Mira_AllLightCurves} presents the same for $o$ Ceti. Samples of the photometry required to produce these light curves are presented in Appendix~\ref{Appendix:LC_Data}. The full tables are given online as Vizier tables.

The $850~\micron$ light curves in general show less scatter than the $450~\micron$ ones. Table \ref{Table:FluxSummary} presents the average flux, standard deviation and the flux range derived from PSF photometry for each source at each wavelength. The average fluxes of both sources at both wavelengths are consistent with the respective total fluxes reported in \citet{Dharmawardena2018}. 

\subsection{Sub-mm Periods}

When determining the periods for both sources, we have to take into account that our light curves are comprised of irregularly sampled data and select a method which will be able to handle these irregularities. We also want to avoid making any prior assumptions of the underlying shape of the periodicity. Therefore we primarily focus on two non-parametric methods to determine our periods. 

For comparison we also employ two commonly used parametric codes, based on Lomb-Scargle (LS) periodograms \citep{Lomb1976-LombScarP, Scargle1982-LombScarP}. LS periodograms are a Fourier transform, power spectrum technique which detects periodicity in a set of time series data which is irregularly sampled. They assume an underlying sine shape for the periodicity. The Lomb-Scargle packages used are the python packages \texttt{Astropy Lomb-Scargle Periodograms} and \texttt{gatspy LombScargle} \citep{VanderPlas2015-GatsPyPaper, GatsPyCode}.

The first non-parametric method we chose is a linear regression algorithm known as Supersmoother, first presented by \citet{Reimann1994_Supersmoother}. We use the implementation in the python package \texttt{gatspy} \citep{VanderPlas2015-GatsPyPaper, GatsPyCode}. Supersmoother is a non-parametric adaptive smoothing algorithm based on a local linear regression estimator with adaptive bandwidths (number of data points).

In essence this algorithm smooths the data along a smoothing length (the range of periods provided as prior estimates, which is then converted to frequencies for calculation purposes in the code). For each candidate frequency within the provided range, the algorithm phase folds the data and fits a straight line to calculate the scatter (density distribution) of the data points. This is carried out repeatedly for all periods in the input range. The period where the scatter is minimised is reported as the optimum period for that data set. This is ideal as the least scatter (scatter minimised the best) should occur when the data is smoothed over exactly one period. 

As the second non-parametric method we chose the python package \texttt{P4J} \citep{Huijse2018_P4J}, which is ideal for unevenly sampled time series data \citep{Principe2010_QMIbook}. The package employs information-theoretic criteria which incorporate information on the whole probability density function of the process and hence is more robust than classical second-order statistics based criteria. It is also not limited to linear relationships or Gaussian probability density functions (PDFs).

The \texttt{P4J} package is based on a type of information criterion known as Quadratic mutual information criterion, where the divergence (statistical distance) between two sets of probability density functions (PDFs) of a random sample is measured: i) the joint PDF of a random sample (PDF of the total sample); and ii) a product of the marginal PDFs (PDFs of each data point in the sample). While there are several methods with which the divergence could measured in \texttt{P4J} we choose to measure the divergence of the PDFs by the the Cauchy–Schwarz statistical inequality (QMICS) in this work. The divergence between the PDFs is checked for a range of periods provided as input priors and the period which gives rise to the least divergent set of PDFs is reported as the best period for that data set. If the light curve is folded with the wrong period, the fluxes are independent of the phases and the structure in the joint PDF will be almost equal to the product of the marginal PDFs. The best-fit period is therefore the one with the maximum correlation between fluxes and phases \citep{Huijse2018_P4J}.

The uncertainties on our final periods were initially determined using Monte Carlo (MC) bootstrapping with one thousand iterations ($\sigma_{MC}$). These uncertainties are comparable to those derived in literature \citep[e.g.,][]{Groenewegen2012IRC+10216}. However, $\sigma_{MC}$ is only a representation of the variations on the location of the peak of the periodograms from independent variations of each data point. As we can see, our periodograms have a wide peak representing a larger range of possible periods and therefore the power can not be localised as well as $\sigma_{MC}$ implies. Therefore we choose to derive another uncertainty, $\sigma_{FWHM}$, which is the FWHM of the periodogram peak. We then present the confidence intervals ($\sigma_{Tot}$) of our period which is the sqaure root of the quadrature sum of $\sigma_{MC}$ and $\sigma_{FWHM}$ as the final uncertainty period. As expected, $\sigma_{Tot}$ is dominated by $\sigma_{FWHM}$ since the FWHMs are much larger compared to the uncertainties derived from MC.

The calculated periods along with their corresponding confidence intervals are given in Tables~\ref{Table:Periods_CWLeo} and ~\ref{Table:Periods_Mira} for CW Leo and Mira respectively. The periodograms and phase folded light curves are presented in appendix \ref{Fig:Periodograms+PhaseCuvres}.

\begin{table*}
  \centering
  \caption{Sub-mm Periods and Confidence Intervals of IRC+10216}
  \begin{threeparttable}
    \begin{tabular}{lllllllll}
    \hline
    \hline
    \multirow{2}[0]{*}{Period Method} & \multicolumn{4}{c}{$450~\micron$ Period (days)} & 
    \multicolumn{4}{c}{$850~\micron$ Period (days)}
   
    \\
   
   & Period & $\sigma_{MC}$ & $\sigma_{FWHM}$ & $\sigma_{Tot}$ & Period & $\sigma_{MC}$ & $\sigma_{FWHM}$ & $\sigma_{Tot}$
   
    \\
    
    \hline

    Astropy LombScargle   & 973 & 1    & 150  & 150 & 678 & 1 & 85 & 85  \\

    Gatspy SlowLomb    & 968 &  6  & 190 & 190 & 668 & 5 & 78 & 79  \\
    
    Gatspy Supersmoother   & 725 &  1  & 80 & 80 & 674 & 5 & 95 & 95 \\
    
    P4J QMICS    & 730 &   5    & 80 & 80 & 667 & 9 & 80 & 80  \\

    \hline
    \hline
    \end{tabular}%
    
    \begin{tablenotes}
    \item The $\sigma_{Tot}$ presented for each wavelength is the square root of the quadrature sum of the $\sigma_{MC}$ and $\sigma_{FWHM}$. We assume $\sigma_{Tot}$ to be the final confidence intervals for our periods. The confidence interval is dominated by the $\sigma_{FWHM}$ given its much larger values. 
    \end{tablenotes}
    
    \end{threeparttable}
   \label{Table:Periods_CWLeo}
\end{table*}%
\begin{table*}
  \centering
  \caption{Sub-mm Periods and Confidence Intervals of \textit{o} Ceti}
  \begin{threeparttable}
    \begin{tabular}{lllllllll}
    \hline
    \hline
    \multirow{2}[0]{*}{Period Method} & \multicolumn{4}{c}{$450~\micron$ Period (days)} & 
    \multicolumn{4}{c}{$850~\micron$ Period (days)}
   
    \\
   
   & Period & $\sigma_{MC}$ & $\sigma_{FWHM}$ & $\sigma_{Tot}$ & Period & $\sigma_{MC}$ & $\sigma_{FWHM}$ & $\sigma_{Tot}$
   
    \\
    
    \hline

    Astropy LombScargle   & 336 & 5    & 29  & 29 & 336 & 1 & 28 & 28  \\

    Gatspy SlowLomb    & 332 &  1  & 30 & 30 & 332 & 2 & 30 & 30  \\
    
    Gatspy Supersmoother   & 350 &  1  & 30 & 30 & 310 & 8 & 95 & 95 \\
    
    P4J QMICS    & 345 &   2    & 31 & 31 & 328 & 3 & 29 & 29  \\

    \hline
    \hline
    \end{tabular}%
    
    \begin{tablenotes}
    \item See table notes in Table~\ref{Table:Periods_CWLeo}. 
    \end{tablenotes}
    
    \end{threeparttable}
   \label{Table:Periods_Mira}
\end{table*}%

\section{Discussion}
\label{Sec:Discussion}

\subsection{Periodograms}

All the periodograms presented show multiple features of different strengths. The most-likely period nearly always dominates over these multiple features having the highest power. We also observe harmonics in the form of somewhat strong peaks appearing at periods of $1/n$ \citep{VanderPlas2018-UnderstandingLombScarP}. The strongest of these harmonics is observed as a strong narrow peak (with a similar but slightly smaller power to the period peak) in the case of IRC+10216 $450~\micron$ between a period of 200 -- 400 days.

In the case of IRC+10216 we find the best fit period peak to be wide with a secondary peak at a lower power blended into it. For the parametric methods at $450~\micron$ this peak appears to correspond to the period determined at $850~\micron$. 

\begin{figure}
  \centering
  \includegraphics[width=0.4\textwidth]{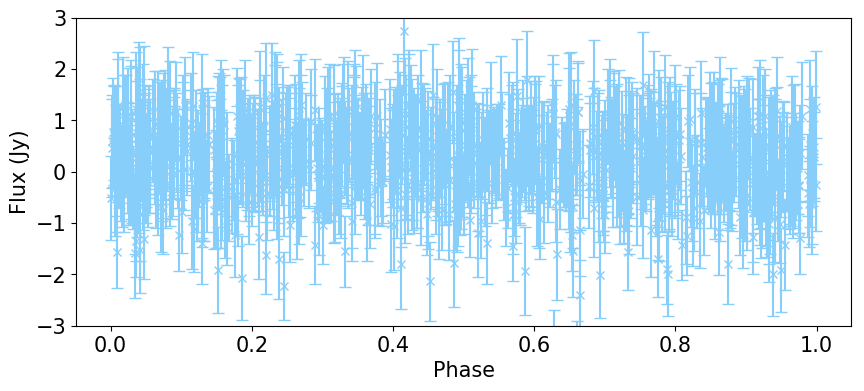}
  \caption{Phase folded residual profile dominated by noise.}
  \label{fig:ResProfile}
\end{figure}

In order to investigate this secondary peak we subtracted the best fit model produced using Astropy Lombscargle from the $850~\micron$ light curve of IRC+10216 and then refitted the data. This does not result in a meaningful period as this light curve (i.e. a residual profile) is dominated entirely by noise resonating around zero flux as shown in Fig.~\ref{fig:ResProfile}. This shows that for $850~\micron$ data a single period describes the data well, at the level of noise seen, as shown by the residual at 850 micron when when one subtracts the strongest period.

Therefore it is likely that the $450~\micron$ IRC+10216 data are noise dominated and hence affecting its period result. Phase folding the $450~\micron$ data with the $850~\micron$ period does not alleviate the scatter observed showing no discernible shape. This further indicates that the $450~\micron$ data is noise dominated. The $850~\micron$ data has a higher signal-to-noise ratio and therefore is preferred in this instance to study the periodicity. 

Further investigation shows that this secondary peak is very likely a result of aliasing with the window function\footnote{The window function is a selection function which describes the observing times of the data set. The light curve in essence is a convolution of the real light curve and this function. The window function can introduce a frequency which can interfere with the true frequency similar to the beat effect}. Using the observing window and the corresponding $850~\micron$ period -- which we assume to be the true period -- we are able to derive a similar period (937 days) to that of the secondary peak, demonstrating this. The aliasing is exacerbated by the noise which dominates the $450~\micron$ data causing the secondary peak be stronger than the real period in some cases. Non parametric methods are better able to handle these adverse effects; the periods determined by these methods at $450~\micron$ are less-strongly affected. Therefore for the periods determined by the parametric methods at $450~\micron$, where the aliased peak dominates we can consider the corresponding $850~\micron$ period to be its real period.


\subsection{Periods and Phase folded Light Curves}

Despite, its light curve being less evenly sampled, the sub-mm periods of $o$ Ceti at both SCUBA-2 wavelengths are much better constrained than those of IRC+10216. There is very good agreement between the periods found at both wavelengths and with the periods reported in the literature \citep{Templeton2009-MiraVariability}. 

The $850~\micron$ period of IRC+10216 are similarly constrained for all methods and is consistent with the optical -- far-IR periods reported in literature \citep{LeBertre1992-CstarLightCurves, Groenewegen2012IRC+10216, Drake2014-CatlinaSurvey, Kim2015-CWLeoVariability}. However at $450~\micron$, while consistent within confidence limits, the periods are significantly larger than those determined at $850~\micron$ and in literature. This effect is likely due a combination of several factors: (i) the uneven sampling of the light curve, (ii) data scatter and (iii) the large uncertainties in individual photometric points. Given the ability of the non-parametric methods (\texttt{P4J} and \texttt{Gatspy Supersmoother}) to better handle uneven sampling, more noisy data and large data scatter we see that these $450~\micron$ periods derived using these two methods are much better constrained compared to the parametric methods. 

We must also note that we are somewhat limited in our ability to constrain the period. The sub-mm data available to us only pertains to $\sim 3$ cycles, given the periods determined. This therefore calls for further sub-mm monitoring of these sources to better estimate and understand the periodicity of these two sources. 

The $850~\micron$ phase folded light curves of IRC+10216 are the least scattered and best sampled, therefore showing a shape to its periodicity in the phase folded light curves. This allows us to determine properties of the periodicity such as phase at which peak brightness occurs. We are unable to repeat this process for the $450~\micron$ and $o$ Ceti counterparts since no such visible shape is present in the phase folded light curves. Once again this most likely the result of the above mentioned factors: data scatter, noisiness and sampling.  

Once we rebin the $850~\micron$ data of IRC+10216, phase fold it using the \texttt{Astropy} LombScargle period\footnote{This period was chosen as it is the largest of the four periods derived for IRC+10216 at $850~\micron$. However the difference between the four periods is only 11 days and therefore the choice of period would not affect our analysis since their differences are at the $1\%$ level.} (678 days) and aligning them to the peak (T0) of the Catalina optical light curve \citep{Drake2014-CatlinaSurvey, Kim2015-CWLeoVariability}, we find that there is a phase difference of $\Delta\phi = \sim 0.79$. This corresponds to a phase lag of $540$ days between the optical light curve and the sub-mm peak of the phase folded IRC+10216 $850~\micron$ light curve.


Further investigation into the phases of IRC+10216 light curves show that from the optical - far-IR the phases are aligned to similar T0 \citep{Drake2014-CatlinaSurvey, Groenewegen2012IRC+10216, Kim2015-CWLeoVariability}. Conversely when we compare the parameters of the radio light curve derived by \citet{Menten2006_CWLeoRadio} to our results at $850~\micron$ we find it to be similar in phase to our data. The peak phase of our data is only $\Delta\phi = \sim 0.2$ off phase with the radio data, given the uncertainties on period and phase determination this slight off-set can be negligible. The period determined by \citet{Menten2006_CWLeoRadio} ($535 \pm 51$ days) is shorter and derived using data covering only $\sim 1$ cycle. However their period is consistent with ours at the $2\sigma$ level. This similarity could point to a related mechanism giving rise to variability at these long wavelengths.

In the following section we explore several possible scenarios which could give rise to the sub-mm variability and the optical -- sub-mm phase lag we observe.

\section{Exploring the Origins the Sub-mm Variability and Phase-lag in IRC+10216}
\label{Sec:OrginOf_Submm}

\subsection{Light Travel Time}
We first investigate whether this phase lag occurs as a result of the light travel times. Optical emission would peak at the moment of the stellar pulse, when the star is at its brightest. The corresponding travel time is $\sim 50$ days 
\citep[0.04 pc;][]{Dharmawardena2018}, too short to explain the phase lag.

\citet{Groenewegen2012IRC+10216} identify a lag of $402 \pm 37$ days between the bow shock and the central source of IRC+10216 in the far-IR, which they attribute to light travel time between the star and the bow shock. This time lag is similar to what we derive. However the bow shock is located at $\sim 3.5'$, well beyond the point source observing window of our SCUBA-2 observations. Therefore we do not detect the bow shock in our sub-mm observations \citep[see][and Sec.\ref{sec:Obs+Reduction} in this work]{Dharmawardena2018}, and it can hence be ruled out as a cause for the sub-mm variability and phase lag.

\subsection{Molecular Line Contamination}
We must also consider the possibility that some of the sub-mm variability and hence the phase lag could be a result of molecular-line contamination. Many publications in the past have shown molecular-line variability of IRC+10216 in great detail \citep{Teyssier2015-CWLeoMolecularLines, He2016-VaraibilityofM33Miras, Fonfria2018_CWLeo_MolLines_3, Pardo2018_CWLeo_MolLines_1}. The variability of these molecular lines may contribute to the variations we observe here and attribute to the (dust) continuum.

\begin{figure}
  \centering
  \includegraphics[width=0.45\textwidth]{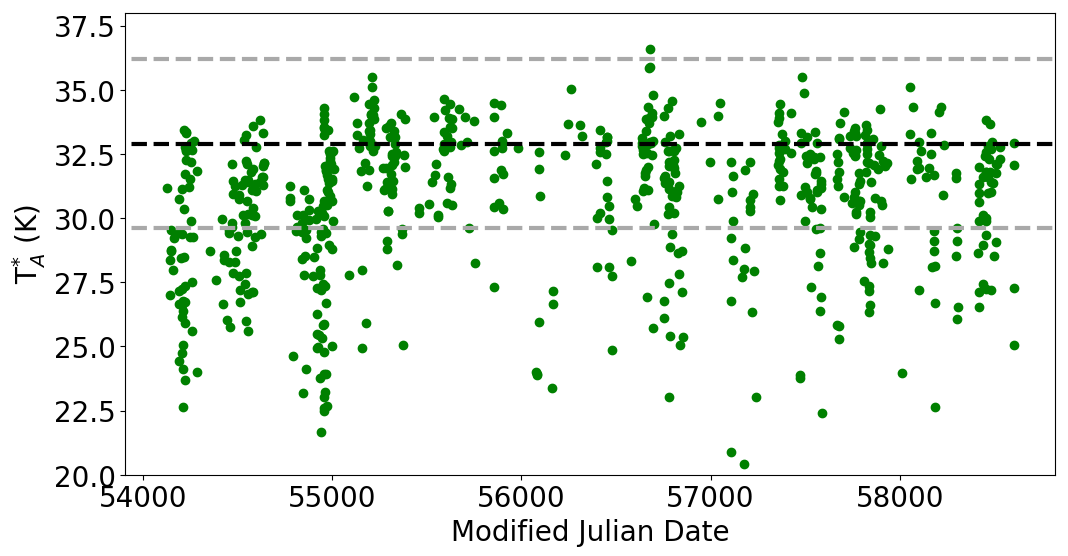}
  \caption{$^{12}$CO(3--2) time series data for IRC+10216 obtained as part of the JCMT/HARP calibration (data and original figure is available in the following JCMT web page: \url{https://www.eaobservatory.org/jcmt/instrumentation/heterodyne/calibration/harp-standards/}). Green dots: time series data; Black  dashed line: The peak antenna temperature of the averaged spectrum of IRC+10216 - 32.9 K; Grey dashed lines: $\pm 10\%$ of the peak antenna temperature of the averaged spectrum.} 
  \label{fig:CO_Monitoring}
\end{figure}

In the case of SCUBA-2 the strongest line in the bandpass and hence the most likely source of line contamination is the $^{12}$CO(3--2) line; other lines are typically much weaker \citep{Drabek2012_SCUBA2_COcontam}. The JCMT has analysed the $^{12}$CO(3--2) line emission of IRC+10216 extensively as part of their calibration process for the HARP instrument\footnote{\url{https://www.eaobservatory.org/jcmt/instrumentation/heterodyne/calibration/harp-standards/}}. In Fig.~\ref{fig:CO_Monitoring} we present the  $^{12}$CO(3--2) time series for IRC+10216 obtained by the JCMT from January 2007 to April 2019.

From figure~\ref{fig:CO_Monitoring} we see that the $^{12}$CO(3--2) line does not appear to show periodic variability at even a $10\%$ level. Attempts made to fit this data to derive a period shows the light curve is dominated by noise and gives rise to an approximately one year period, which likely corresponds to the annual gaps in the data when the source is not visible. In addition, the $^{12}$CO(3--2) line typically contributes only about $\sim 8-20\%$ to the SCUBA-2 $850~\micron$ flux \citep{Drabek2012_SCUBA2_COcontam, Parsons2018_SCUBA2_COsubtraction}. In the case of IRC+10216, assuming a line width of 30 km s$^{-1}$ and an average CO flux of 32.9 K (from Fig.~\ref{fig:CO_Monitoring}) we find an effective continuum contribution from CO of 0.5 Jy or $5\%$ of the continuum flux. Taking these factors into account the maximum change in the continuum due to $^{12}$CO(3--2) is no more than $0.5\%$ and therefore is not the cause of the sub-mm variability or phase lag observed.

\subsection{Cycle of Dust Formation and Destruction}
We also explore the possibility that the relationship between the stellar pulsation and dust formation cycle may explain the sub-mm variability and observed phase lag. The star's brightness and size increases during the stellar pulse and is best observed in the optical - near-IR, hence giving rise to the optical peak. At the stellar minimum, the material in the surrounding circumstellar envelope is at its coolest and therefore most favourable for dust formation (causing the sub-mm emission to peak). Dust is then illuminated and subsequently heated undergoing thermal emission over time. As the star approaches maximum light some of this dust is sublimated reducing the total dust mass. We suggest this cyclic dust cooling, formation, heating and destruction close to the central star is responsible for the sub-mm variability and hence the lag observed between the optical and sub-mm peaks. 


Support for this provided by the self consistent C-rich AGB wind model \texttt{DARWIN} \citep{Bladh2019_DarwinModels}. This model shows that the dust density increases out of phase in relation to the stellar pulse. The dust condensation region changes by $\sim 1$ R$_{*}$ with more dust formed closer to the central star at minimum stellar luminosity ($\phi = 0.5$) (see their Fig. 2, middle panel).

\subsubsection{Radiative Transfer Modelling of IRC+10216}
We generate radiative transfer models of IRC+10216 using the python package \texttt{Hyperion} \citep{Robitaille2011_Hyperion}. By doing so we explore the origins of the sub-mm variability, phase lag and a simplified version of the above scenario.

With \texttt{Hyperion} we are limited to static models as this method can not handle time-dependent radiative transfer or self consistent dust formation and destruction (unlike the \texttt{DARWIN} wind model). As a result we are only able to approximately model the light curve. The inherent non-linearity of the radiative transfer problem means that interpolation between minimum and maximum luminosity snapshots would not capture the physical process at work. 

Instead we present snapshots built along the bolometric luminosity light curve of IRC+10216 which allows us to generate a snapshot light curve in the sub-mm ignoring the time dependent effects. The stellar parameters were taken from \citet{Menten2006_CWLeoRadio, Menten2012-CWLeo} and the dust parameters are derived by \citet{Dharmawardena2018}. Key stellar parameters used are as follows: (i) Maximum stellar luminosity (L$_{max}$): $1.1\times10^{4}$ L$_{\odot}$; (ii) Minimum stellar luminosity (L$_{min}$): $4400$ L$_{\odot}$; (iii) Effective stellar temperature: 2750 K; (iv) Stellar radius (R$_{*}$): 1.9 AU. In the case of the dust grain properties we assume a $\kappa_{1\micron}^{S} = 1.5\times10^{4}$ cm$^{2}$g$^{-1}$. The dust is assumed to be a mixture of $90\%$ amorphous carbon and $10\%$ silicon carbide with their optical constants taken from \citet{Zubkoetal1996} and \citet{Pegourie1988} respectively. The grain size distribution is assumed to be a power-law with an exponential fall off as determined by \citet{Kim1994_DustProperties}. Here we use a minimum grain size of $0.01~\micron$ along with an exponential cut-off scale of $1~\micron$  and a power-law slope of $-3.5$. \texttt{Hyperion} uses \texttt{BHMIE} to calculate the weighted averaged grain cross section to use as the model input from the grain distribution described.

Hyperion iteratively removes dust above our input dust sublimation temperatures, therefore essentially assuming instant dust formation and destruction. This therefore ignores several parameters dictating dust formation/destruction in AGB stars, e.g., chemistry, shocks and hydrodynamical effects.

\begin{figure}
  \centering
  \includegraphics[width=0.4\textwidth]{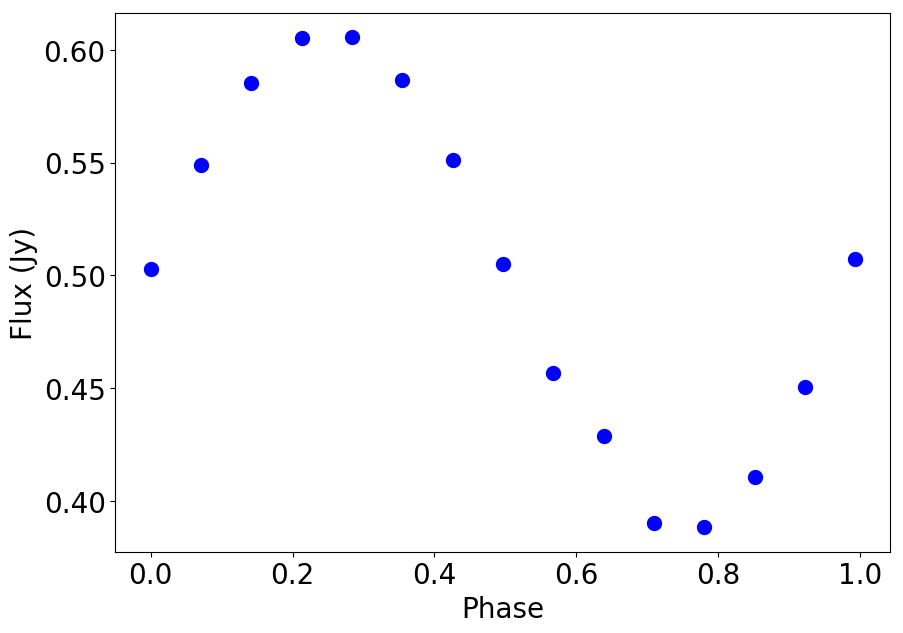}
  \caption{Modelled sub-mm Light curve generated using a series \texttt{Hyperion} static models.}
  \label{fig:Model_Submm_LC}
\end{figure}

The resulting model sub-mm light curve is presented in Fig.~\ref{fig:Model_Submm_LC}. While we observe a variable model curve we do not observe a phase lag compared to the input bolometric luminosity light curve. In order to test any time delays in dust formation/destruction we require time-dependent radiative transfer models such as \texttt{DARWIN}.

Further we only recover $\sim 4\%$ of the observed flux at $850~\micron$ with each model forming the model light curve. However this is not an issue for our analysis as we are interested in the relative changes in flux and the resulting dust parameter, such as the dust condensation radius.

However comparing the peak/trough ratio between the observed $850~\micron$ light curve and the model light curve we find them both to be a similar ratio. The ratio of the observed light curve is $1.4 \pm 0.2$ while the model ratio is $1.6$. Therefore we are able to recover the same magnitude of variation as the observed light curve.

\begin{figure}
  \centering
  \includegraphics[width=0.4\textwidth]{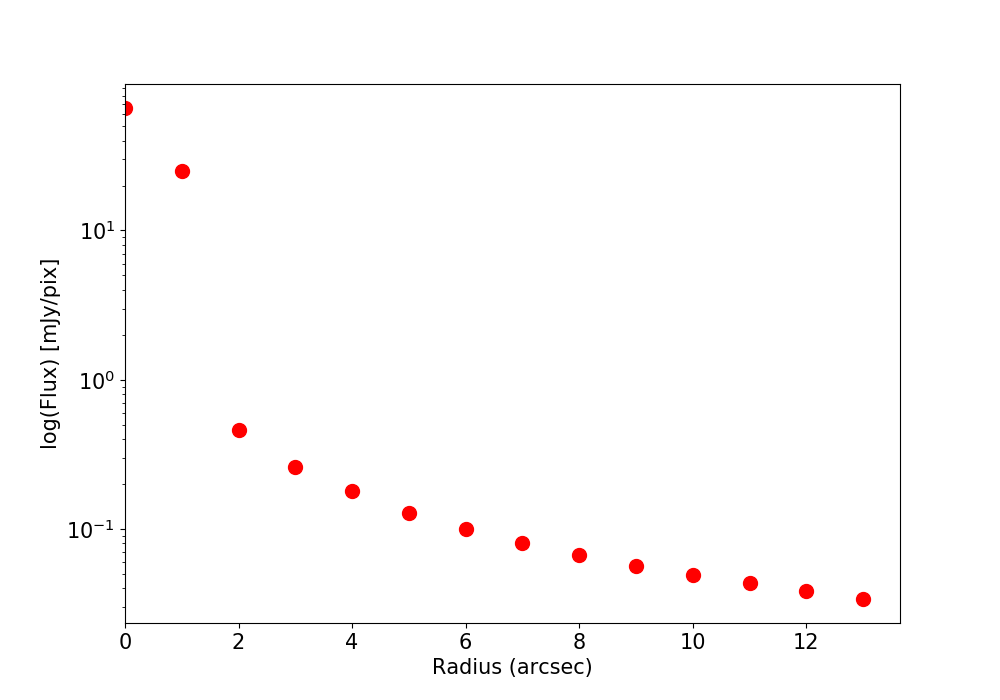}
  \caption{Radial profile derived from peak sub-mm flux model depicted up to the SCUBA-2 850~$\micron$ beam FWHM ($13\arcsec$).} 
  \label{fig:Model_RadProf}
\end{figure}

Moreover, we are able to narrow down the region from which the bulk of the sub-mm emission arises by studying the radial profiles derived from our models (see Fig.~\ref{fig:Model_RadProf}). By studying the radial profiles of our models we find that $99\%$ of the flux within the beam FWHM ($13\arcsec$ at 850$~\micron$) arises from the small inner $2\arcsec$ (260 AU) region. Therefore we are not sensitive to historic thermal dust emission from the outer circumstellar envelope along the line of sight of the beam FWHM in this instance. This inner $2\arcsec$ region well encompasses the expected dust formation region of IRC+10216 as well as the extended radio photosphere described in \citet{Menten2006_CWLeoRadio}.

\begin{figure}
  \centering
  \includegraphics[width=0.45\textwidth]{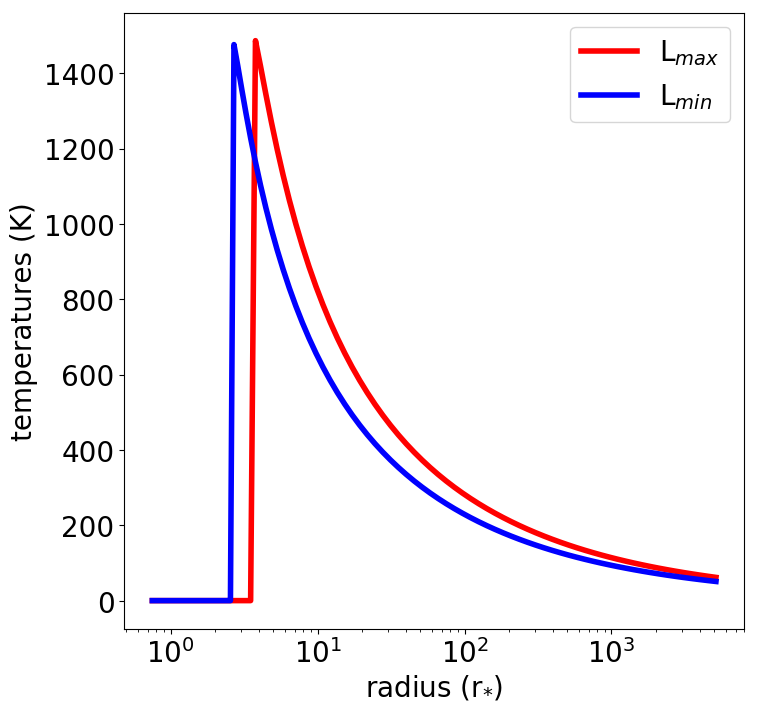}
  \caption{Temperature as a function of stellar radius as a function of maximum stellar luminosity (red; stellar pulse at its peak) and minimum stellar luminosity (blue; stellar pulse at its weakest)}
  \label{fig:HypModel_CWLeo}
\end{figure}

Given a dust sublimation temperature \texttt{Hyperion} has the ability to calculate the dust sublimation region by iteratively removing dust above the sublimation temperature until the radiative transfer converges. We interpret this radius as the dust condensation radius. Using our models we find a shift in dust condensation radii by $\sim 1.0$ R$_{*}$ going from 3.5 R$_{*}$ at L$_{max}$ to 2.5 R$_{*}$ at L$_{min}$ (see Fig.~\ref{fig:HypModel_CWLeo}). This is consistent with the shift in dust condensation radii predicted by the \texttt{DARWIN} wind models.

Therefore looking at our modelled results it is likely that the dust formation and destruction cycle may contribute to the variability of IRC+10216 to a limited degree. However some other mechanism in the inner regions of the envelope dominates over this.

\subsection{Free-free emission}

As discussed above, we find our phase lag to be similar to that observed in the radio by \citet{Menten2006_CWLeoRadio}. \citet{Menten2006_CWLeoRadio, Menten2012-CWLeo} suggest the radio emission likely arises from a radio photosphere $\sim$ twice the stellar radius. This photosphere extends approximately up to the beginning of the dust formation region observed in our models. They further suggest the dominant emission mechanism likely to cause the variability in the radio photosphere is free-free emission. These authors suggest that shocks travelling at speeds of approximately 10 -- 15 km~s$^{-1}$ can heat up the material causing the free-free emission in the outflow and producing the variability at 7mm.

Using the sub-mm - radio levels of free-free emission presented in \citet{Menten2006_CWLeoRadio} (see their Fig. 2) we calculate the largest possible contribution of flux from free-free emission to our variability. At $850~\micron$ free-free emission contributes 950 mJy, extrapolating from 3 mm with the spectral index of 1.96 provided in \citet{Menten2006_CWLeoRadio}. Therefore, we find that the free-free emission contributes only $\sim 10\%$ of the $850~\micron$ flux that we observe. By the same logic applied to molecular lines above this rules free-free emission out as a likely cause for the observed sub-mm variability.

Therefore given the current constraints on our models we are unable to pin-point the cause of the sub-mm emission. However the coupled nature of the sub-mm and radio light curves suggests that both are driven by a single physical cause. Further investigation into this as of yet unknown mechanism is required to see what role it may play in the sub-mm continuum emission.

\section{Conclusions}
\label{Sec:Conclusions}

We characterise the sub-mm periodicity of two of the most well-known AGB stars; IRC+10216 and $o$ Ceti. The observations were obtained from pointing calibrations from the JCMT/SCUBA-2 instrument at $450~\micron$ and $850~\micron$. 

In order to avoid assuming an underlying shape and to better account for the irregular sampling of the light curve we were motivated to use two non-parametric methods for period determination. For comparison we also employed two parametric methods. The derived periods of $o$ Ceti agreed very well with published results and with each other at the two wavelengths. The periods range from 332--350 days at $450~\micron$ and 310 to 336 $850~\micron$, also agreeing well with each others uncertainties. The $450~\micron$ observations of IRC+10216, where the data scatter, large uncertainties in individual data points and the uneven sampling affects the period determination the most, the non-parametric methods were more effective in constraining the period. The periods of IRC+10216 at $850~\micron$ for all four methods agreed well with past publications, ranging from 667 -- 678 days.

Overall the periods determined have large confidence intervals. The precision and confidence interval of these sub-mm periods can in the future be improved using the following methods: 1) reducing the discontinuities in data by continuing to monitor these stars in the sub-mm; 2) reduce the uncertainties on the individual photometric points, and hence the scatter in data points. While higher signal-to-noise individual observations would help improve the uncertainties somewhat, in the $850~\micron$ light curves we can clearly see that they are calibration limited and therefore deeper observations will not be very effective. Instead the individual photometric uncertainties can be improved by determining the relative uncertainties in flux variations using field stars. In order to do so however, we must expand the observation field in order to incorporate field stars in the field of view. The JCMT Transients project successfully uses this method to determine accurate light curves to detect unknown variability \citep{Mairs2017}. 

We observe a $\sim 540$ day time lag between the published optical light curves and our $850~\micron$ IRC+10216 light curve despite having similar periods at all wavelengths. A similar phase lag is observed in the radio, while shorter wavelengths match well with the optical. Studying a set of models in the literature in combination with our own radiative transfer models suggests that there are several mechanisms contribute to the sub-mm variability. The relationship between the stellar pulse, luminosity and the dust formation cycles may play a role, as well as an as yet unknown dominant mechanism which results in both the sub-mm and radio variability. Shocks waves and dust heating may be responsible. Due to the limitations placed on our static models we are unable to discern the nature of this dominant mechanism. Time-dependent self-consistent radiative-hydrodynamical models, high resolution ALMA observations ($< 0.1\arcsec$) and time resolved imaging in the sub-mm will allow us to shed light on this in the future. 

Should the dust formation cycle be found to dominate sub-mm light curves may be a key method for studying the life cycle of dust formation and destruction in the innermost regions of the stellar outflow which is extremely difficult to directly observe without the aid of deep high-resolution observations from facilities as ALMA and VLT.

\section*{Acknowledgements}
We thank the anonymous referee for their insightful response which helped improve the paper. 
We thank Prof. A. Zijlstra for useful discussions which contributed to this manuscript. 
TED wishes to thank Prof. Chung-Ming Ko at NCU for his support of this project.
This research has been supported under grants MOST104-2628-M-001-004-MY3 and MOST107-2119-M-001-031-MY3 from the Ministry of Science and Technology of Taiwan, and grant  AS-IA-106-M03 from Academia Sinica.

The James Clerk Maxwell Telescope is operated by the East Asian Observatory on behalf of The National Astronomical Observatory of Japan; Academia Sinica Institute of Astronomy and Astrophysics; the Korea Astronomy and Space Science Institute; the Operation, Maintenance and Upgrading Fund for Astronomical Telescopes and Facility Instruments, budgeted from the Ministry of Finance (MOF) of China and administrated by the Chinese Academy of Sciences (CAS), as well as the National Key R\&D Program of China (No. 2017YFA0402700). Additional funding support is provided by the Science and Technology Facilities Council of the United Kingdom and participating universities in the United Kingdom and Canada.
In addition to software cited above, this research made use of the \textsc{Scipy} \citep{Scipy2001} and \textsc{Astropy} \citep{Astropy2018} python packages.
This research used the facilities of the Canadian Astronomy Data Centre operated by the National Research Council of Canada with the support of the Canadian Space Agency. This research also made use of the Canadian Advanced Network for Astronomical Research \citep[CANFAR,][]{Gaudet2010-CANFAR}.
We acknowledge with thanks the variable star observations from the AAVSO International Database contributed by observers worldwide and used in this research.

\bibliographystyle{mnras}
\bibliography{Sub-mm_Bib}

\appendix
\section{Derived Periodograms and Light Curves}
\label{Fig:Periodograms+PhaseCuvres}

\begin{figure*}
\centering
\begin{subfigure}[b]{0.3\textwidth}
  \centering
  \includegraphics[width=\textwidth]{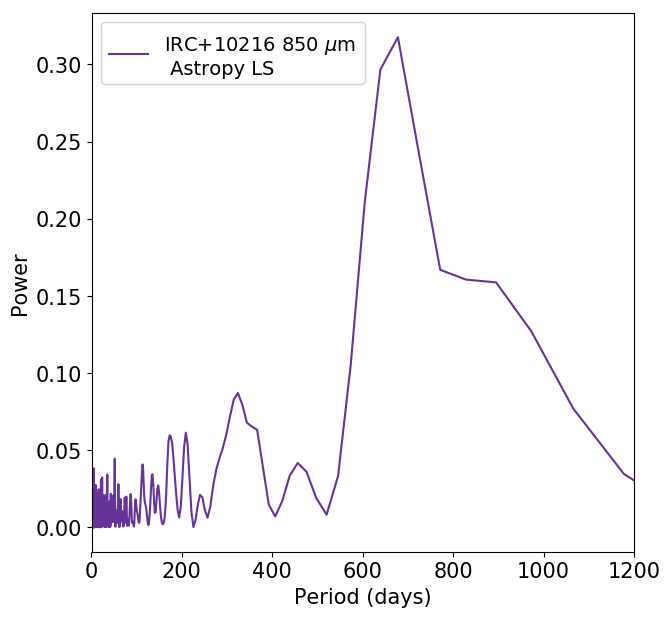}
  \label{fig:CWLeo850Astropy_Periodogram}
  \end{subfigure}
\begin{subfigure}[b]{0.5\textwidth}
  \centering
  \includegraphics[width=\textwidth]{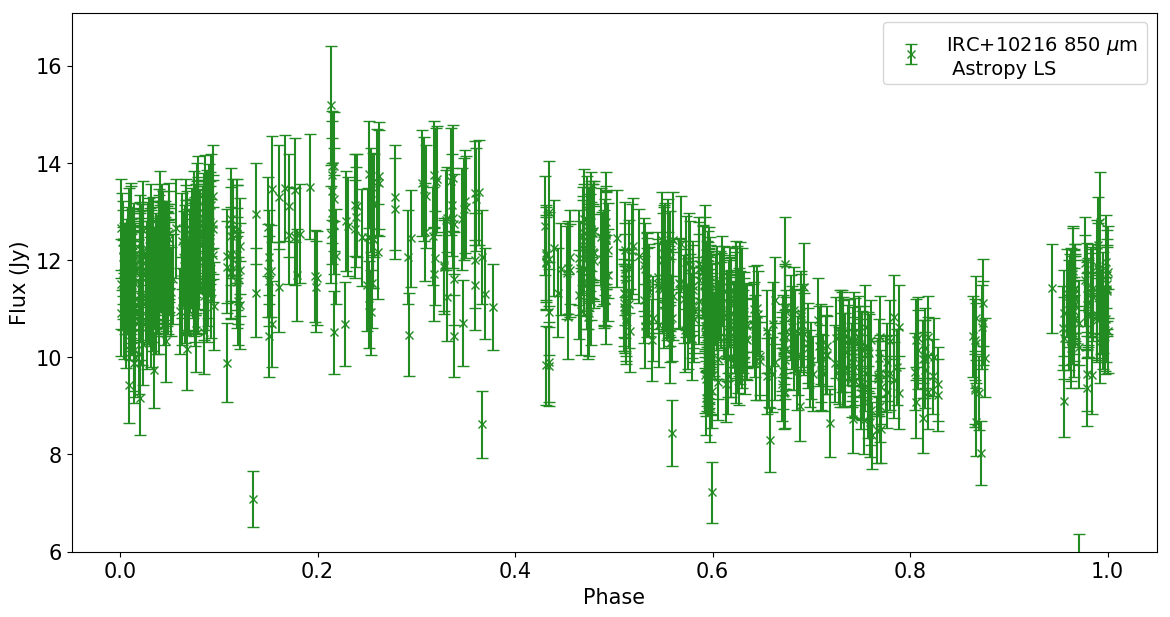}
  \label{fig:CWLeo850Astropy_LightCurve}
  \end{subfigure}
  
\begin{subfigure}[b]{0.3\textwidth}
  \centering
  \includegraphics[width=\textwidth]{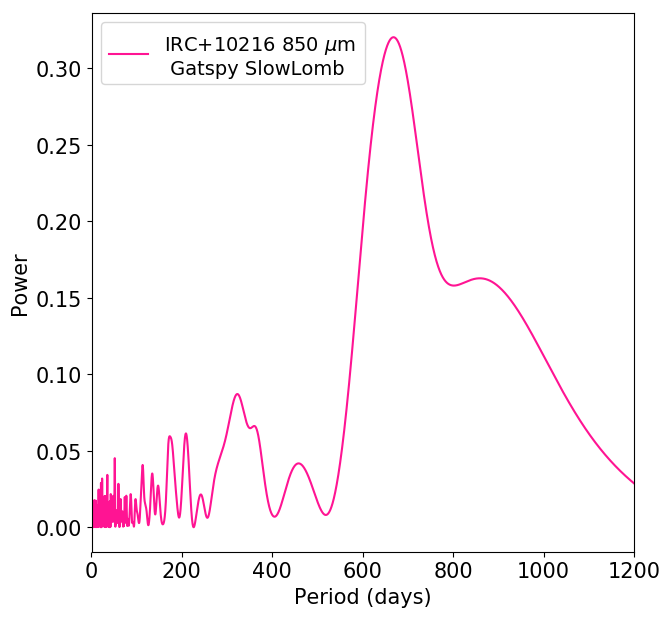}
  \label{fig:CWLeo850GatspySL_Periodogram}
  \end{subfigure}
\begin{subfigure}[b]{0.5\textwidth}
  \centering
  \includegraphics[width=\textwidth]{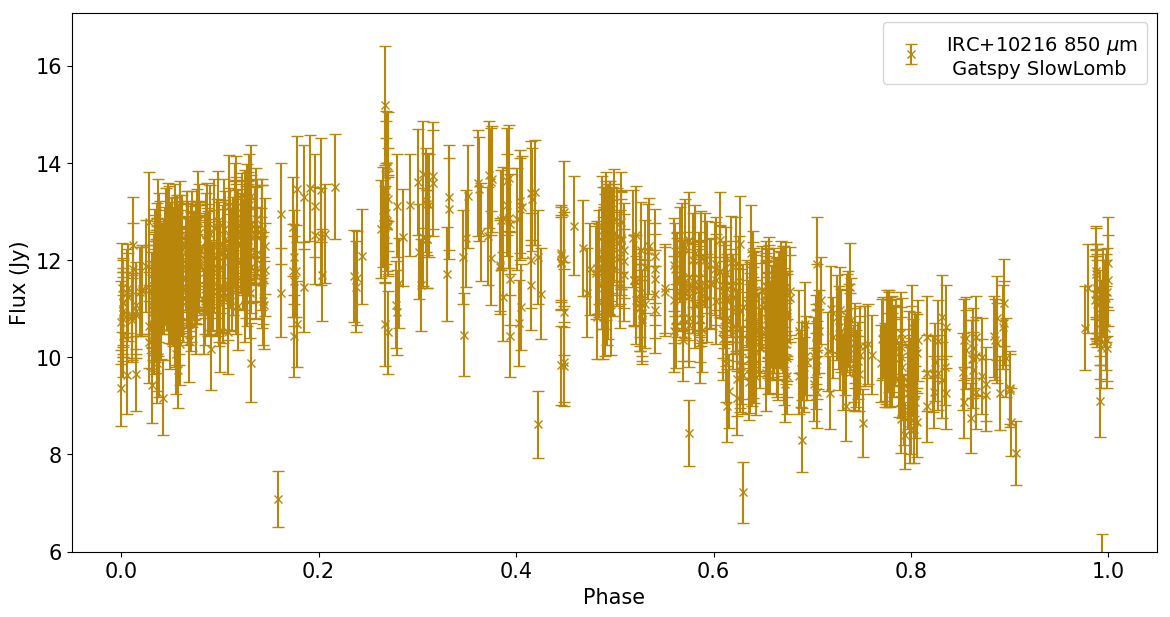}
  \label{fig:CWLeo850GatspySL_LightCurve}
  \end{subfigure}  
  
\begin{subfigure}[b]{0.3\textwidth}
  \centering
  \includegraphics[width=\textwidth]{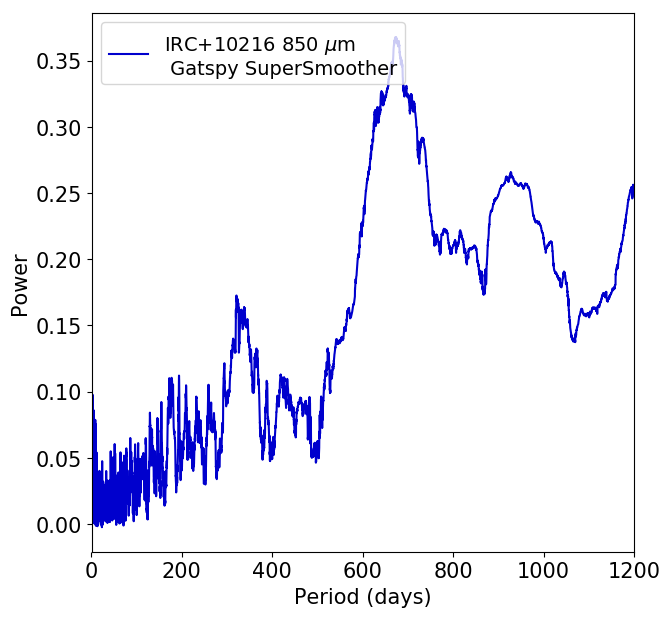}
  \label{fig:CWLeo850GatspySS_Periodogram}
  \end{subfigure}
\begin{subfigure}[b]{0.5\textwidth}
  \centering
  \includegraphics[width=\textwidth]{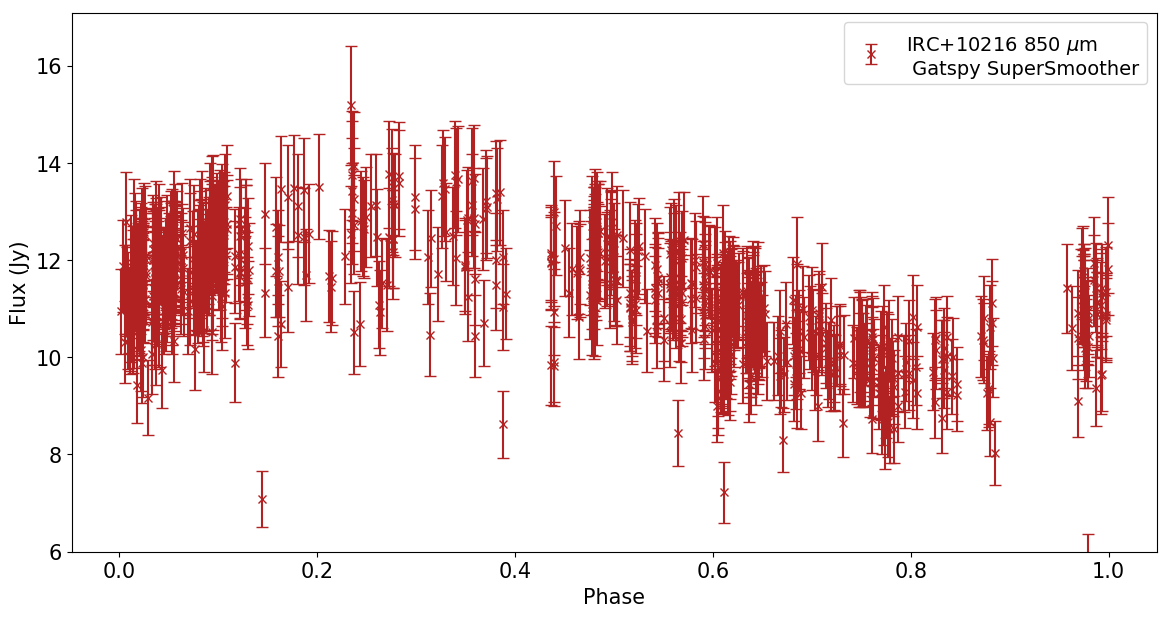}
  \label{fig:CWLeo850GatspySS_LightCurve}
  \end{subfigure}  
  
\begin{subfigure}[b]{0.3\textwidth}
  \centering
  \includegraphics[width=\textwidth]{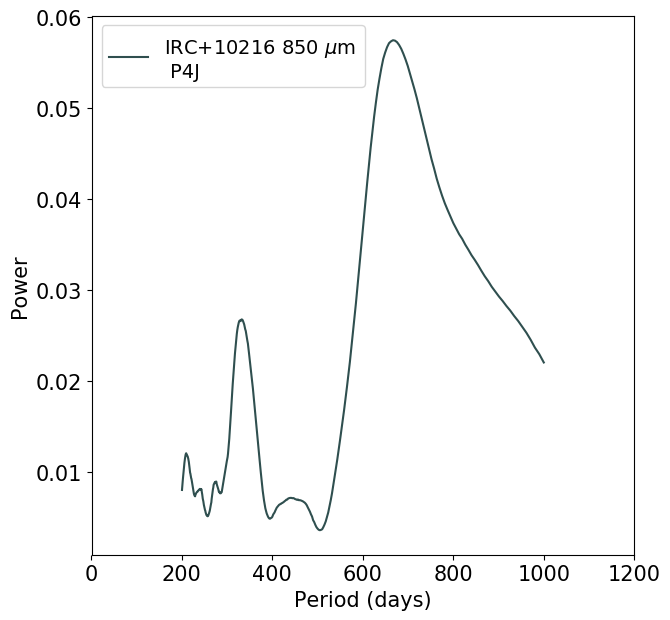}
  \label{fig:CWLeo850P4J_Periodogram}
  \end{subfigure}
\begin{subfigure}[b]{0.5\textwidth}
  \centering
  \includegraphics[width=\textwidth]{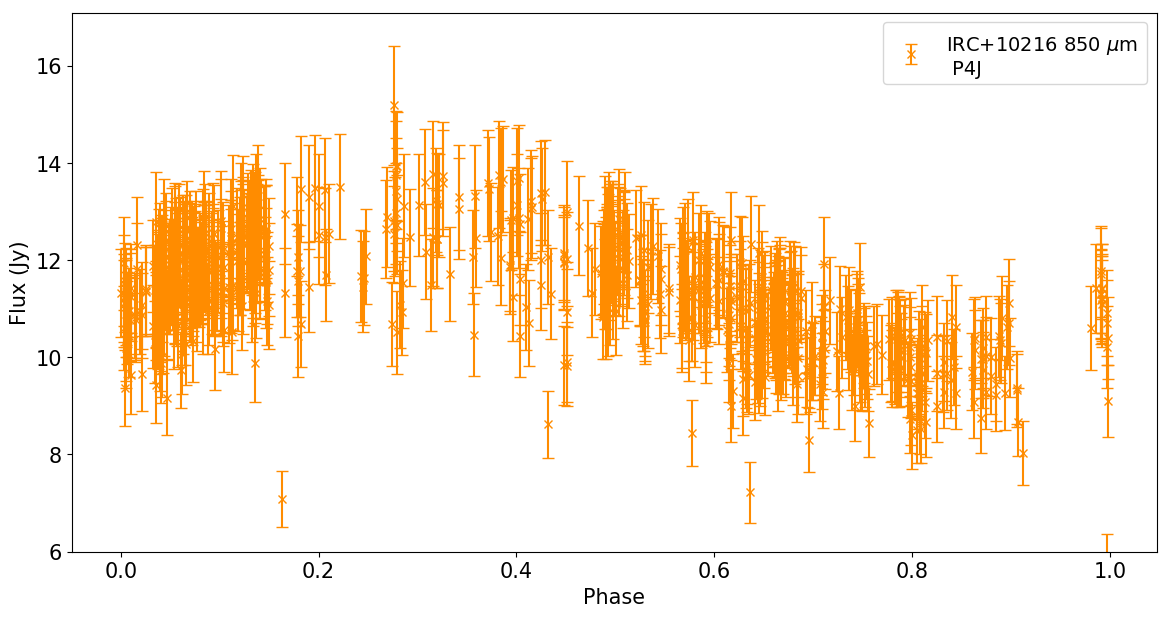}
  \label{fig:CWLeo850P4J_LightCurve}
  \end{subfigure}

  \caption{Derived Periodograms and Phase folded light curves for IRC+10216 $850~\micron$. The light curve is phase folded using the far-IR T0 from \citet{Groenewegen2012IRC+10216} and the peak is shifted by $\delta\phi = 0.45$ to better present the shape of the curve.}
  \label{fig:CWLeo_850_AllCurves}
\end{figure*}

\clearpage
\begin{figure*}
\centering
\begin{subfigure}[b]{0.3\textwidth}
  \centering
  \includegraphics[width=\textwidth]{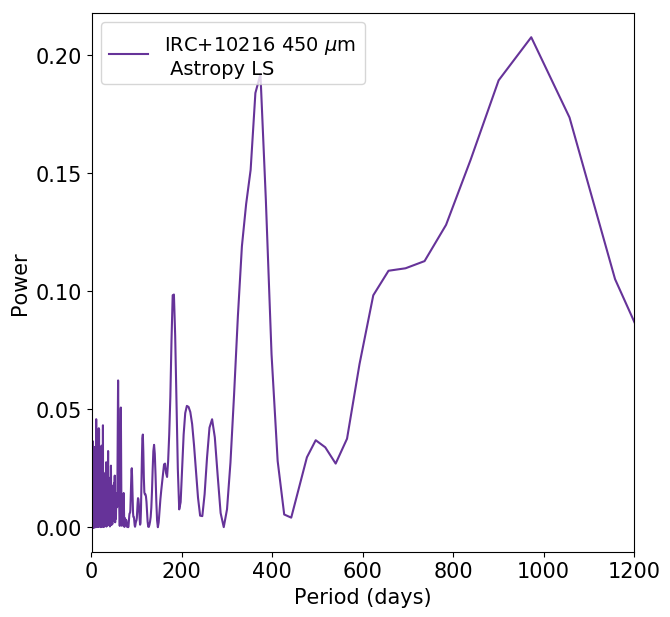}
  \label{fig:CWLeo450Astropy_Periodogram}
  \end{subfigure}
\begin{subfigure}[b]{0.5\textwidth}
  \centering
  \includegraphics[width=\textwidth]{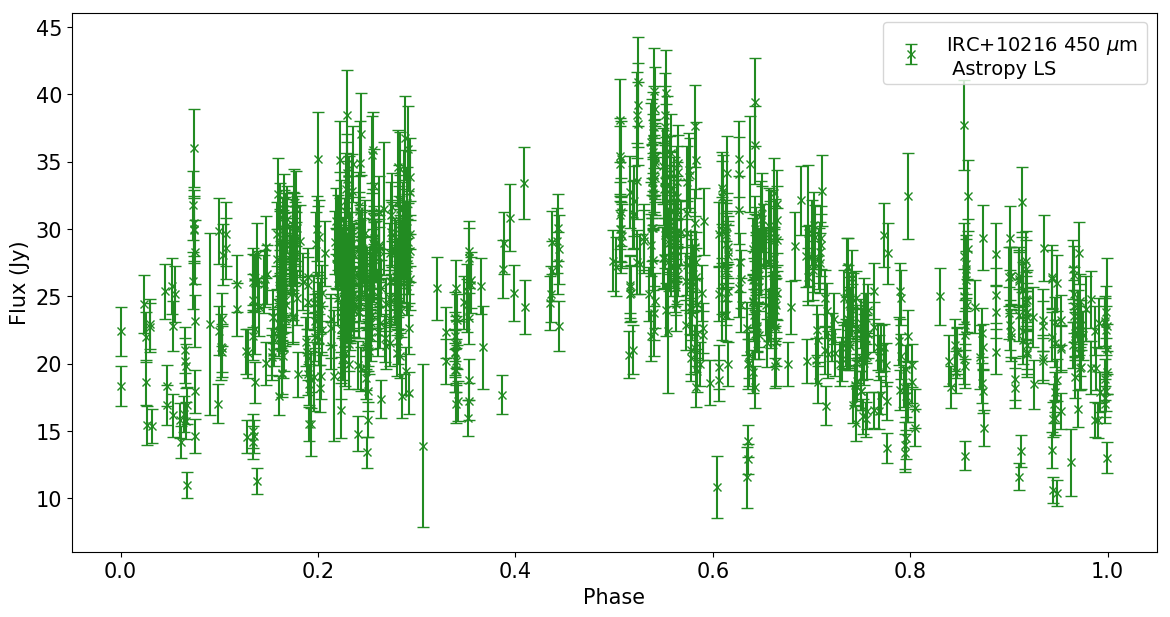}
  \label{fig:CWLeo450Astropy_LightCurve}
  \end{subfigure}
  
\begin{subfigure}[b]{0.3\textwidth}
  \centering
  \includegraphics[width=\textwidth]{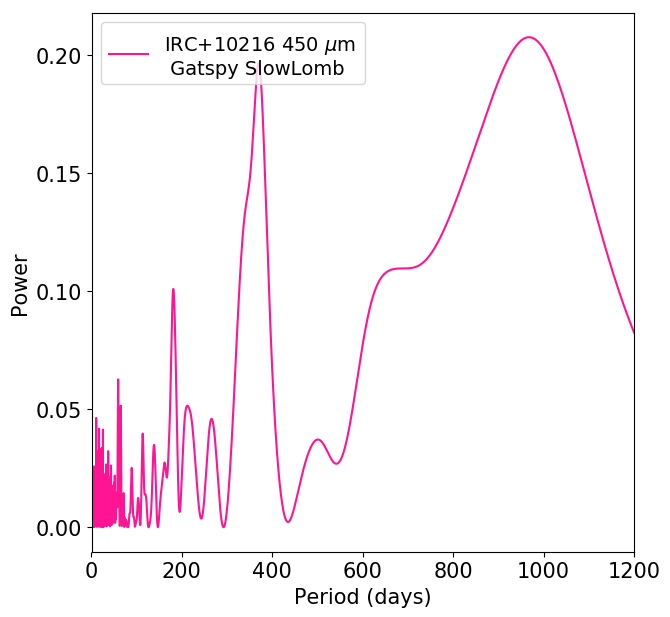}
  \label{fig:CWLeo450GatspySL_Periodogram}
  \end{subfigure}
\begin{subfigure}[b]{0.5\textwidth}
  \centering
  \includegraphics[width=\textwidth]{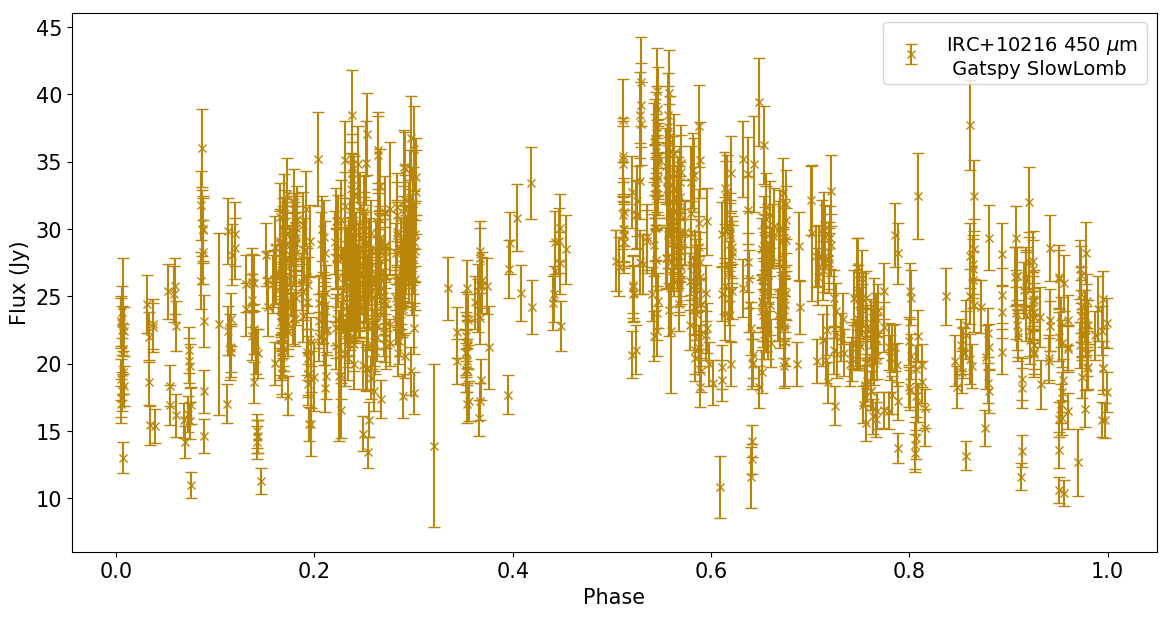}
  \label{fig:CWLeo450GatspySL_LightCurve}
  \end{subfigure}
  
\begin{subfigure}[b]{0.3\textwidth}
  \centering
  \includegraphics[width=\textwidth]{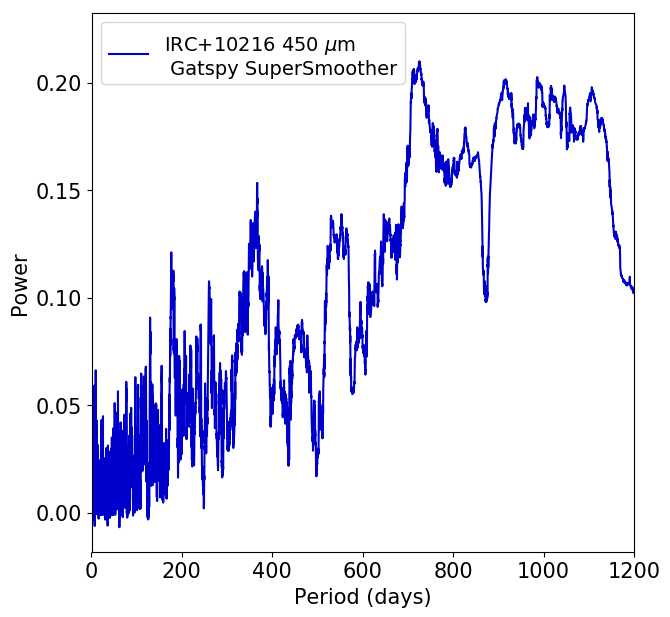}
  \label{fig:CWLeo450GatspySS_Periodogram}
  \end{subfigure}
\begin{subfigure}[b]{0.5\textwidth}
  \centering
  \includegraphics[width=\textwidth]{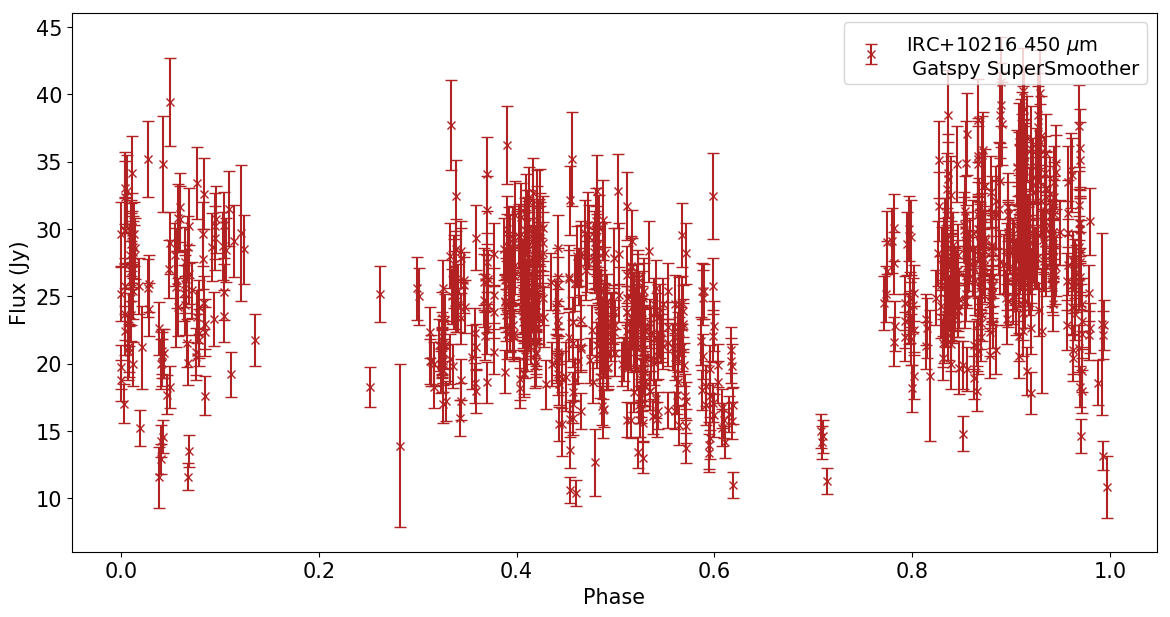}
  \label{fig:CWLeo450GatspySS_LightCurve}
  \end{subfigure}
  
\begin{subfigure}[b]{0.3\textwidth}
  \centering
  \includegraphics[width=\textwidth]{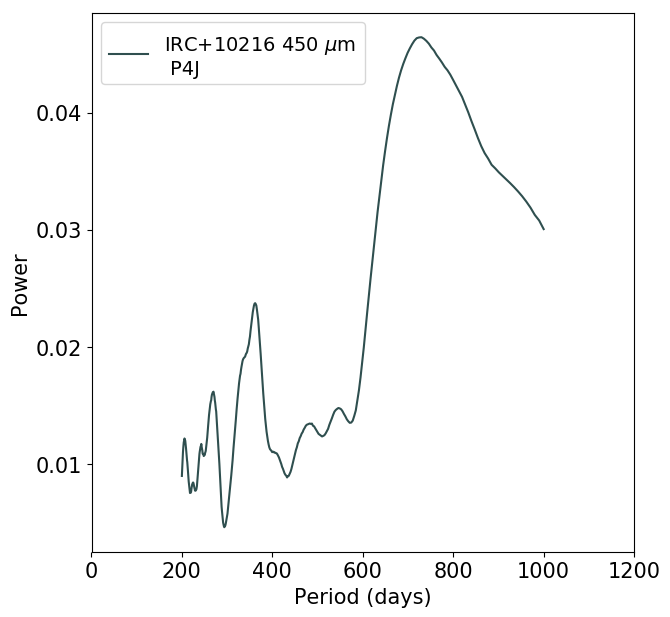}
  \label{fig:CWLeo450P4J_Periodogram}
  \end{subfigure}
\begin{subfigure}[b]{0.5\textwidth}
  \centering
  \includegraphics[width=\textwidth]{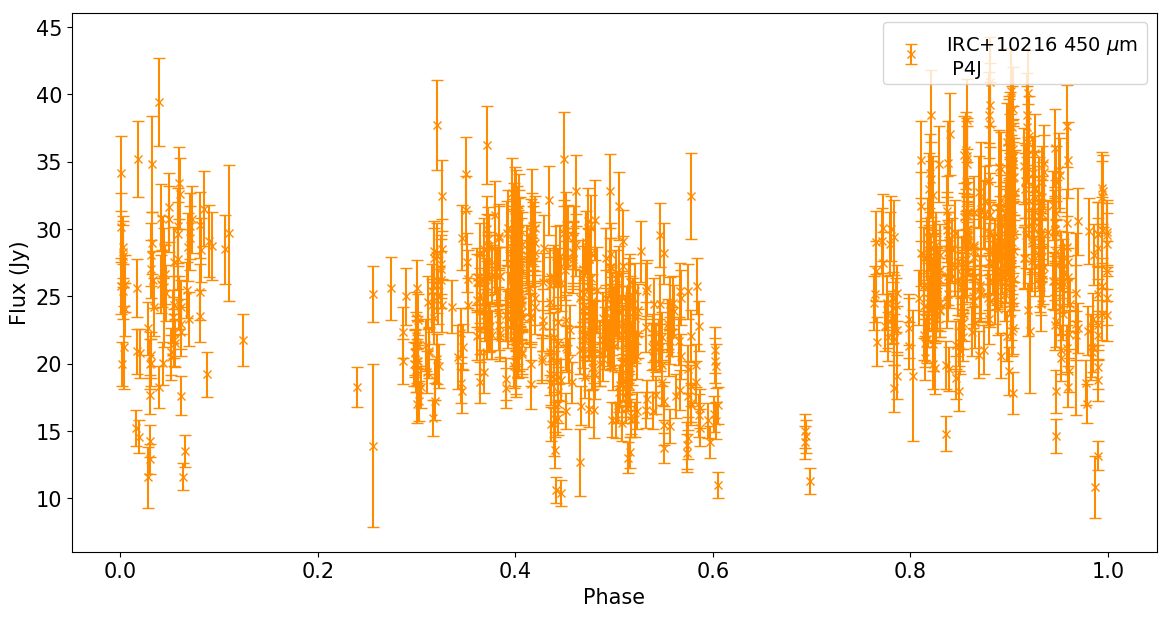}
  \label{fig:CWLeo450P4J_LightCurve}
  \end{subfigure}

  \caption{Same as Fig.~\ref{fig:CWLeo_850_AllCurves}, for IRC+10216 $450~\micron$}
  \label{fig:CWLeo_450_AllCurves}
\end{figure*}

\clearpage
\begin{figure*}
\centering
\begin{subfigure}[b]{0.3\textwidth}
  \centering
  \includegraphics[width=\textwidth]{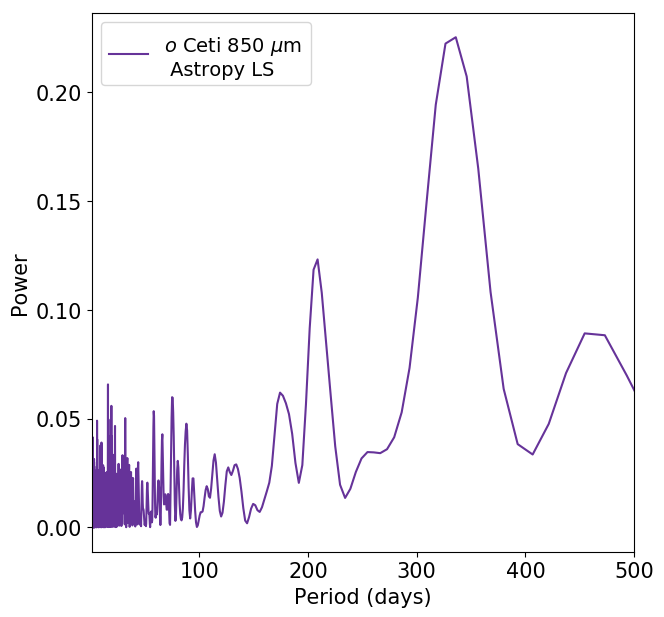}
  \label{fig:Mira850Astropy_Periodogram}
  \end{subfigure}
\begin{subfigure}[b]{0.5\textwidth}
  \centering
  \includegraphics[width=\textwidth]{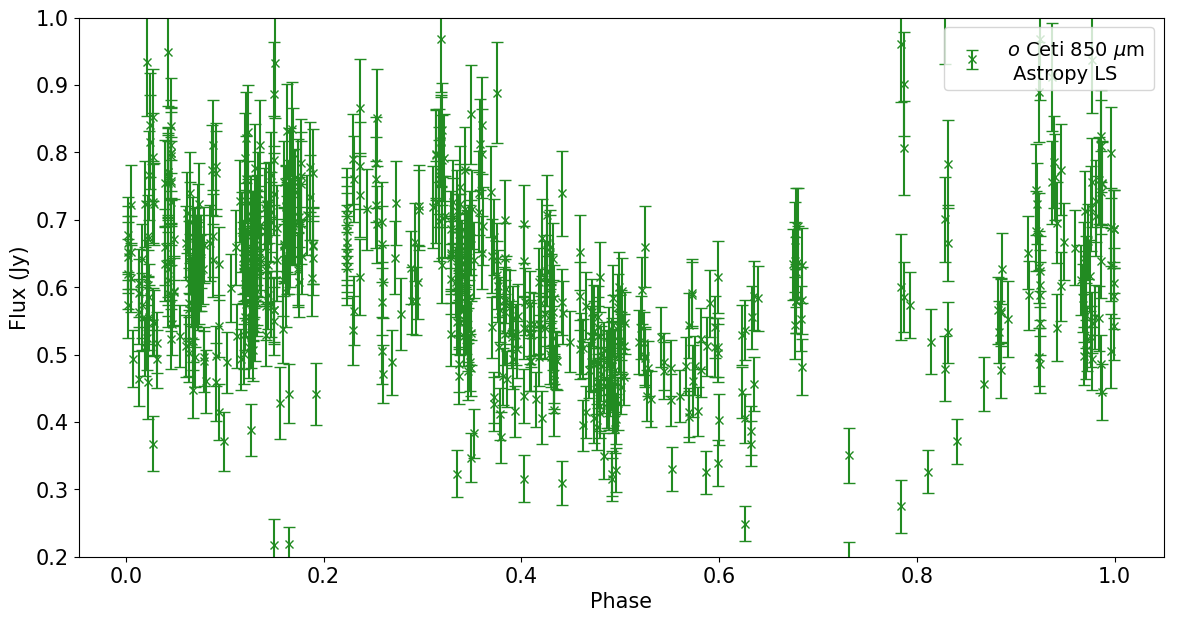}
  \label{fig:Mira850Astropy_LightCurve}
  \end{subfigure}
  
\begin{subfigure}[b]{0.3\textwidth}
  \centering
  \includegraphics[width=\textwidth]{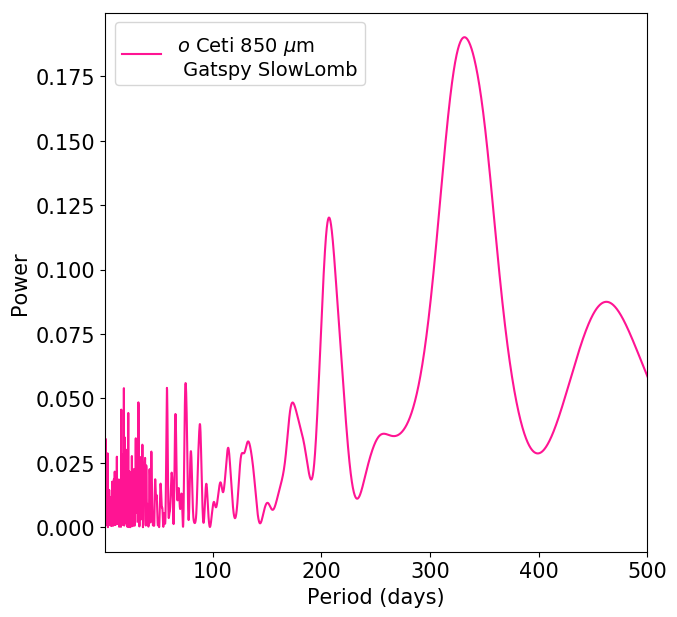}
  \label{fig:Mira850GatspySL_Periodogram}
  \end{subfigure}
\begin{subfigure}[b]{0.5\textwidth}
  \centering
  \includegraphics[width=\textwidth]{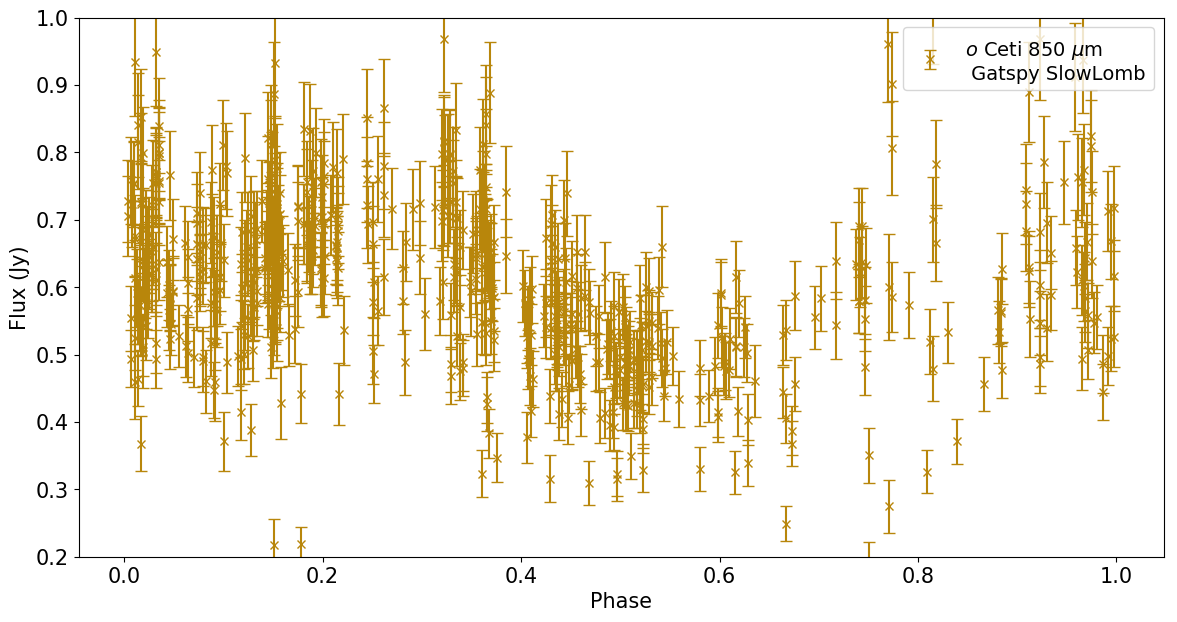}
  \label{fig:Mira850GatspySL_LightCurve}
  \end{subfigure}  
  
\begin{subfigure}[b]{0.3\textwidth}
  \centering
  \includegraphics[width=\textwidth]{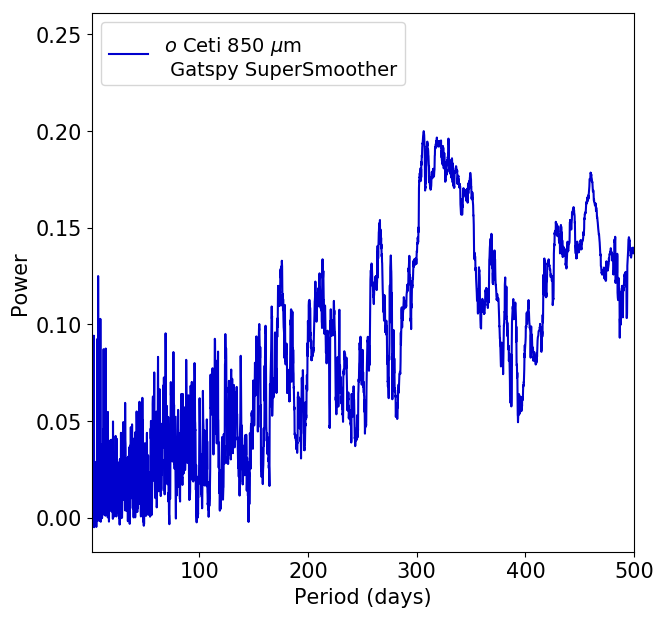}
  \label{fig:Mira850GatspySS_Periodogram}
  \end{subfigure}
\begin{subfigure}[b]{0.5\textwidth}
  \centering
  \includegraphics[width=\textwidth]{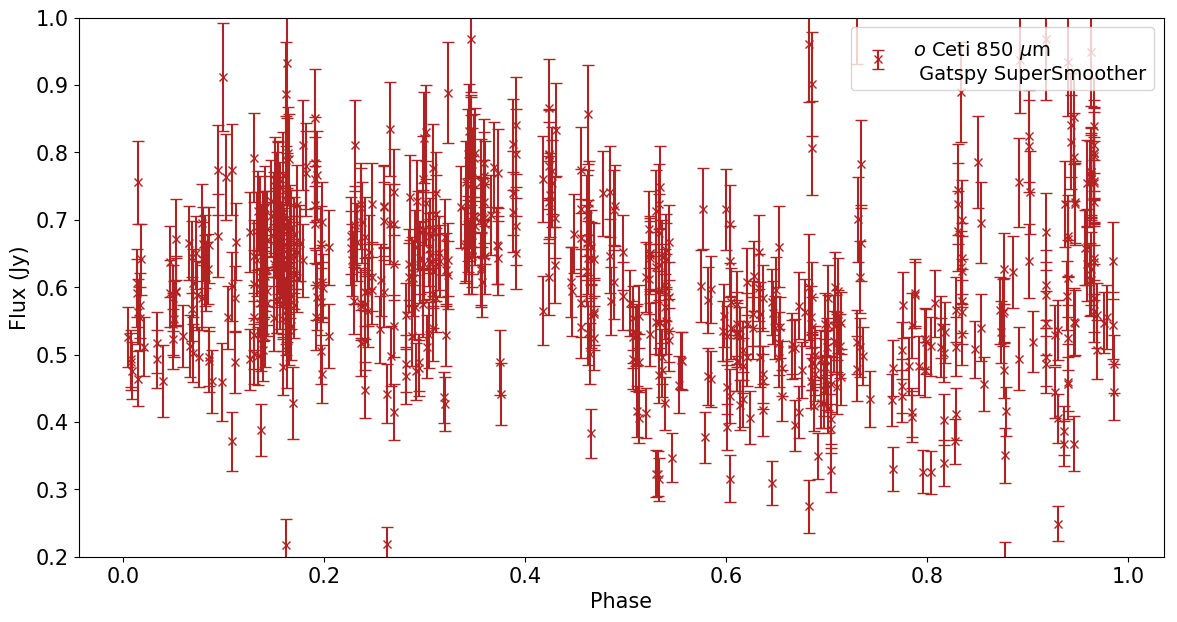}
  \label{fig:Mira850GatspySS_LightCurve}
  \end{subfigure}  
  
\begin{subfigure}[b]{0.3\textwidth}
  \centering
  \includegraphics[width=\textwidth]{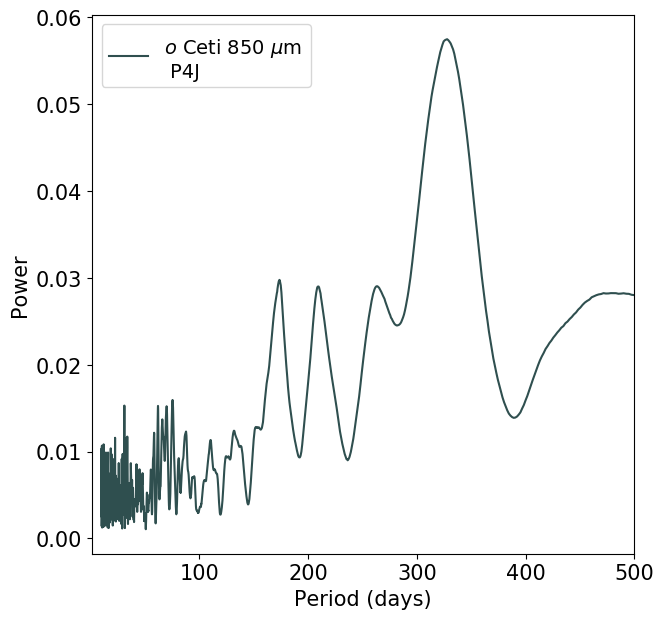}
  \label{fig:Mira850P4J_Periodogram}
  \end{subfigure}
\begin{subfigure}[b]{0.5\textwidth}
  \centering
  \includegraphics[width=\textwidth]{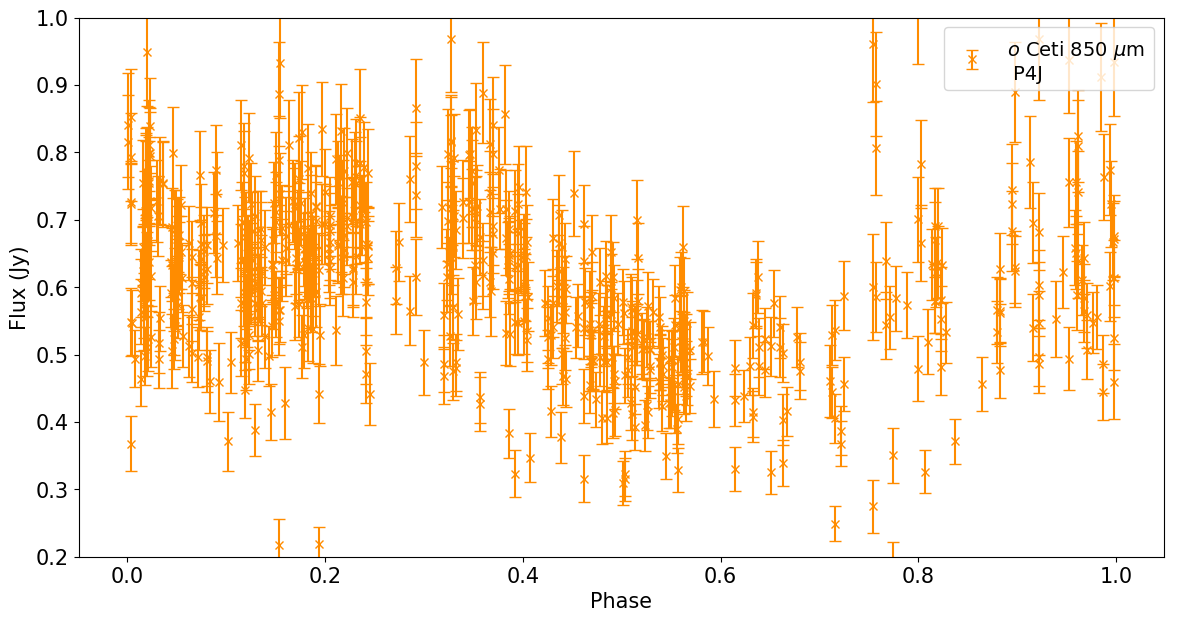}
  \label{fig:Mira850P4J_LightCurve}
  \end{subfigure}

  \caption{Derived Periodograms and Phase folded light curves for $o$ Ceti $850~\micron$. The data is phase folded to the peak of the AAVSO visual and V band light curves.}
  \label{fig:Mira_850_AllCurves}
\end{figure*}

\clearpage
\begin{figure*}
\centering
\begin{subfigure}[b]{0.3\textwidth}
  \centering
  \includegraphics[width=\textwidth]{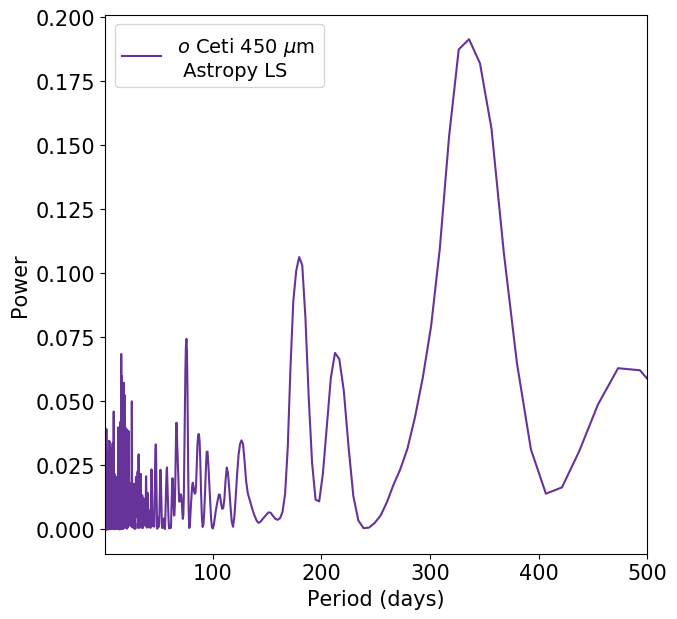}
  \label{fig:Mira450Astropy_Periodogram}
  \end{subfigure}
\begin{subfigure}[b]{0.5\textwidth}
  \centering
  \includegraphics[width=\textwidth]{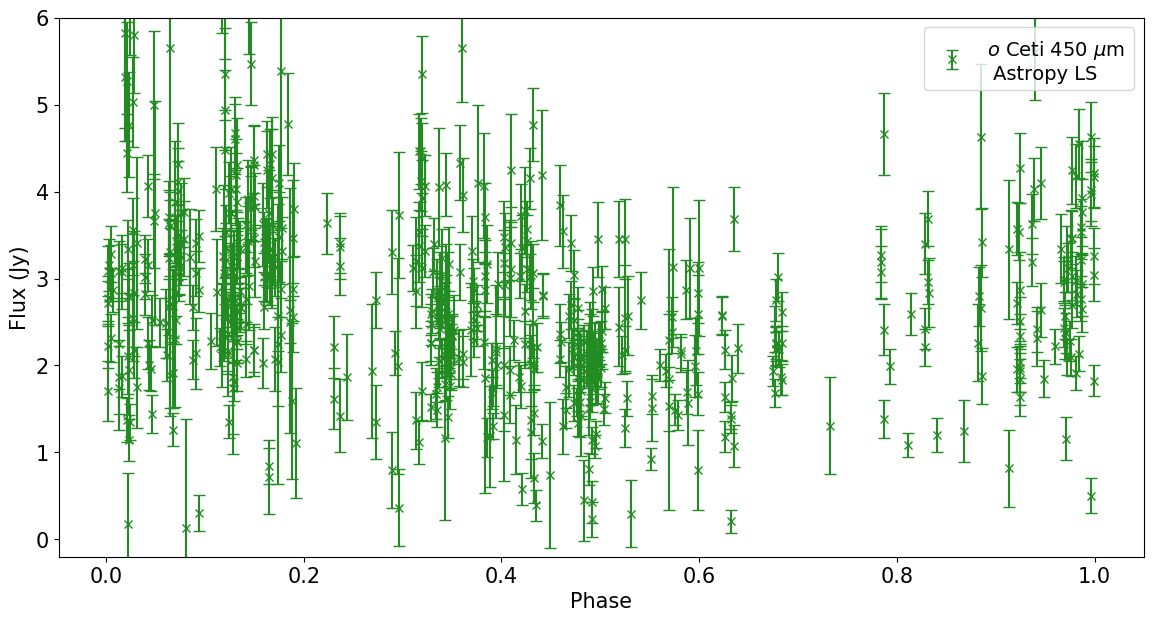}
  \label{fig:Mira450Astropy_LightCurve}
  \end{subfigure}
  
\begin{subfigure}[b]{0.3\textwidth}
  \centering
  \includegraphics[width=\textwidth]{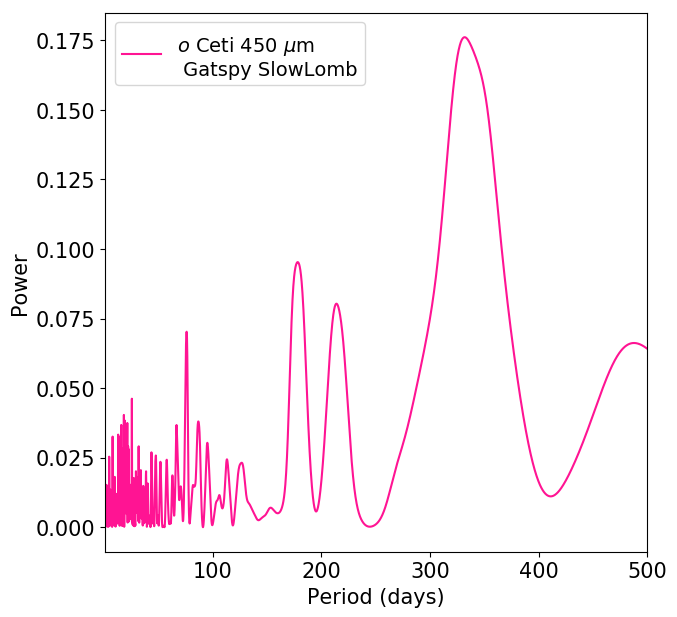}
  \label{fig:Mira450GatspySL_Periodogram}
  \end{subfigure}
\begin{subfigure}[b]{0.5\textwidth}
  \centering
  \includegraphics[width=\textwidth]{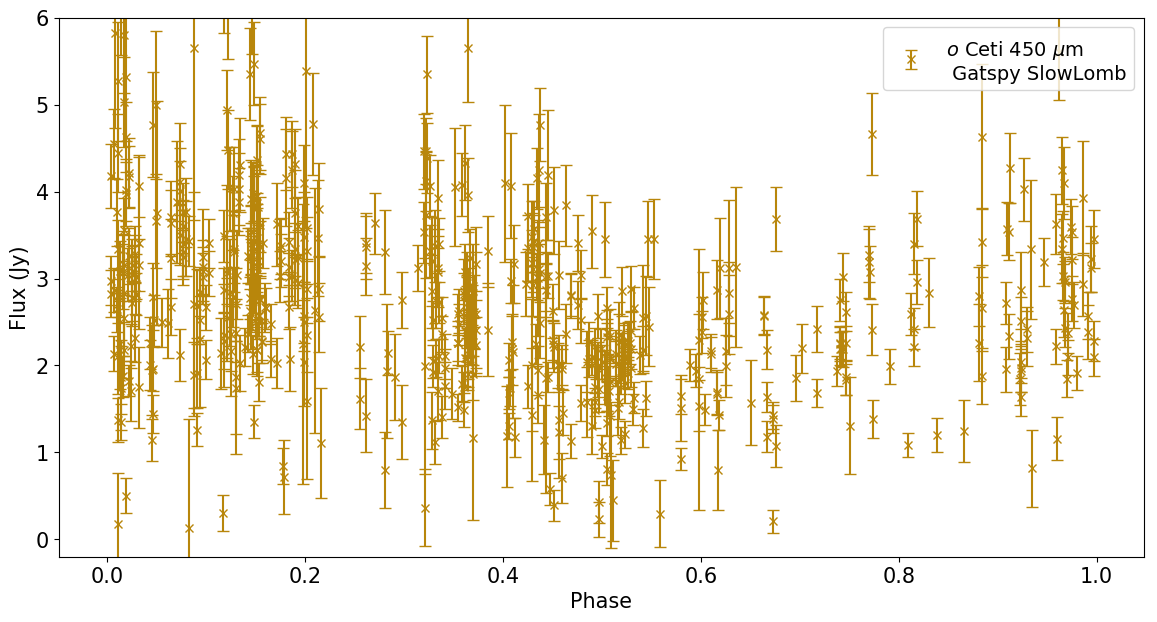}
  \label{fig:Mira450GatspySL_LightCurve}
  \end{subfigure}
  
\begin{subfigure}[b]{0.3\textwidth}
  \centering
  \includegraphics[width=\textwidth]{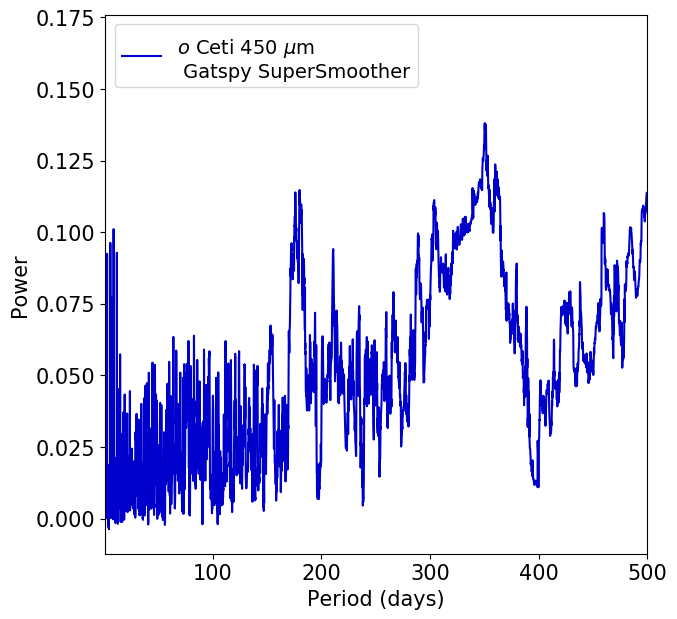}
  \label{fig:Mira450GatspySS_Periodogram}
  \end{subfigure}
\begin{subfigure}[b]{0.5\textwidth}
  \centering
  \includegraphics[width=\textwidth]{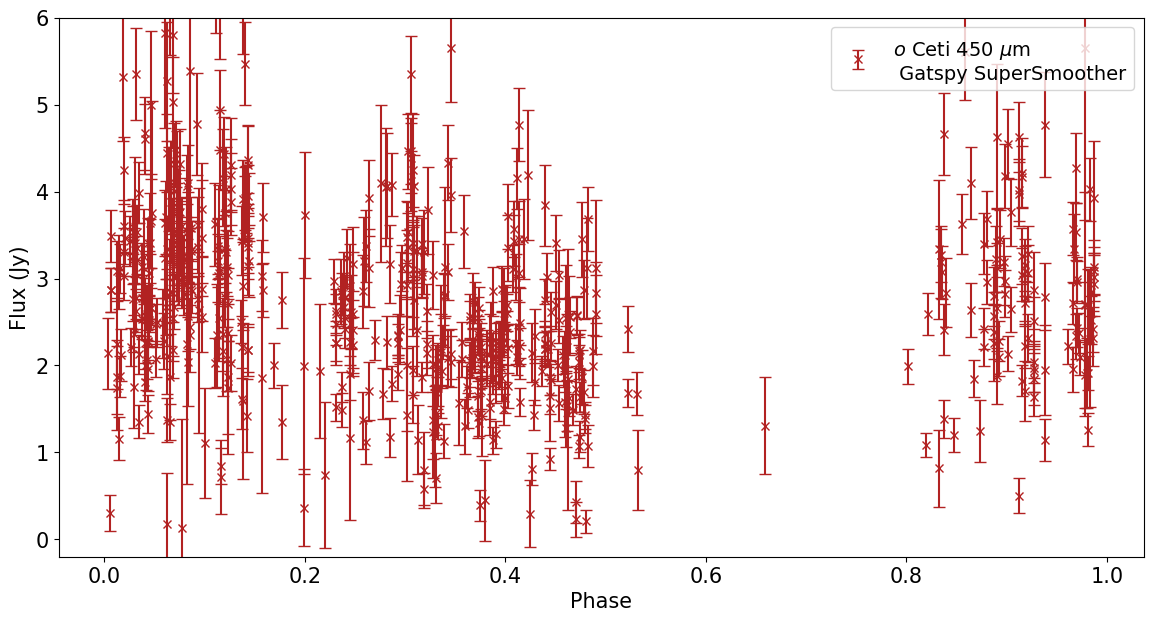}
  \label{fig:Mira450GatspySS_LightCurve}
  \end{subfigure}
  
\begin{subfigure}[b]{0.3\textwidth}
  \centering
  \includegraphics[width=\textwidth]{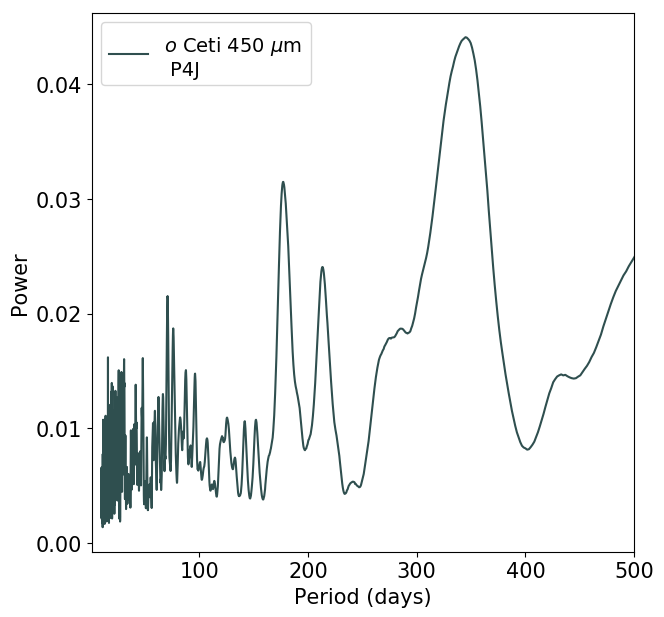}
  \label{fig:Mira450P4J_Periodogram}
  \end{subfigure}
\begin{subfigure}[b]{0.5\textwidth}
  \centering
  \includegraphics[width=\textwidth]{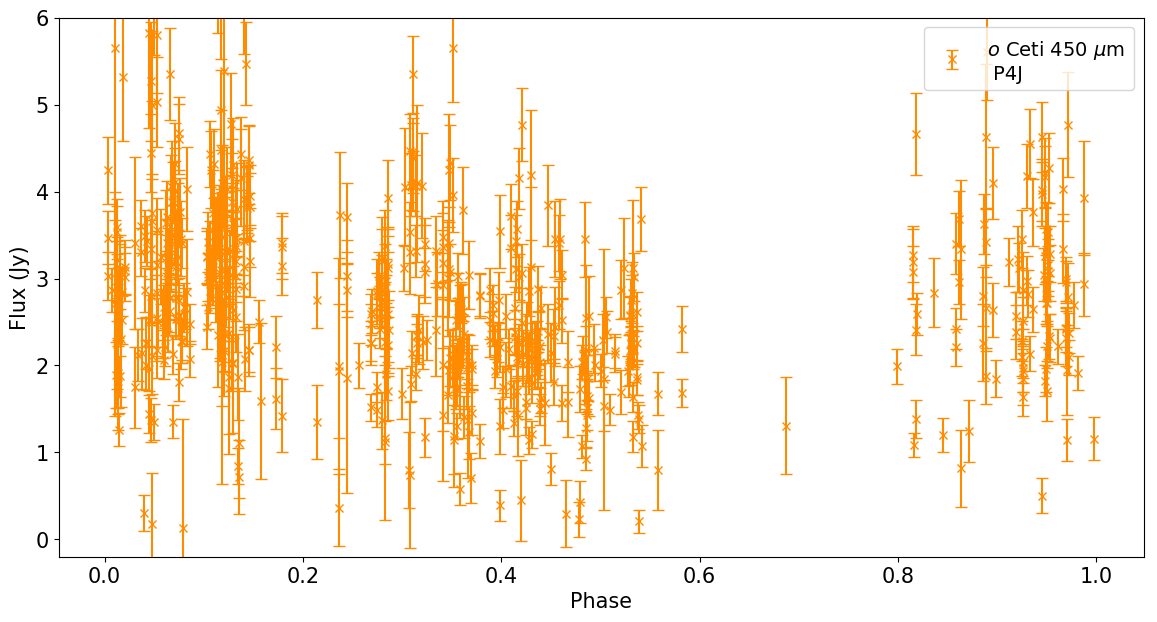}
  \label{fig:Mira450P4J_LightCurve}
  \end{subfigure}

  \caption{Same as Fig.~\ref{fig:Mira_850_AllCurves}, for $o$ Ceti $450~\micron$}
  \label{fig:Mira_450_AllCurves}
\end{figure*}

\newpage
\clearpage

\section{Samples of Source Light Curve Data}
\label{Appendix:LC_Data}

A sample of the data required to produce the light curves presented in Figs.~\ref{fig:CWLeo_AllLightCurves} and ~\ref{fig:Mira_AllLightCurves} is presented here. The full tables are given online as vizier tables.

\begin{table}
  \centering
  \caption{Sample of IRC+10216 450$\micron$ Light Curve Data}
    \begin{tabular}{lll}
    
    \hline
    \hline
    
    MJD & Flux & $\sigma_{F}$ \\
    &  (Jy) & (Jy)\\
    
    \hline
    
55650	&	22.8	&	1.84	\\
55650	&	22.2	&	1.78	\\
55650	&	13.2	&	1.10	\\
55669	&	15.2	&	1.32	\\
55704	&	11.6	&	1.02	\\
55705	&	13.5	&	1.19	\\
55707	&	25.5	&	2.07	\\
55845	&	25.2	&	2.09	\\
55921	&	26.5	&	2.12	\\
55921	&	26.1	&	2.09	\\

    \hline
    \hline
    
    \end{tabular}%
  \label{Table:CWLeo_LCdata_450}%
\end{table}%

\begin{table}
  \centering
  \caption{Sample of $o$ Ceti 450$\micron$ Light Curve Data}
    \begin{tabular}{lll}
    
    \hline
    \hline
    
    MJD & Flux & $\sigma_{F}$ \\
    &  (Jy) & (Jy)\\
    
    \hline
    
55764	&	3.19	&	0.41	\\
55764	&	3.27	&	0.30	\\
55764	&	3.07	&	0.31	\\
55765	&	4.67	&	0.47	\\
55765	&	2.41	&	0.29	\\
55765	&	1.38	&	0.22	\\
55778	&	3.40	&	0.35	\\
55779	&	2.42	&	0.24	\\
55779	&	2.21	&	0.21	\\
55780	&	3.69	&	0.32	\\

    \hline
    \hline
    
    \end{tabular}%
  \label{Table:Mira_LCdata_450}%
\end{table}%

\begin{table}
  \centering
  \caption{Sample of IRC+10216 850$\micron$ Light Curve Data}
    \begin{tabular}{lll}
    
    \hline
    \hline
    
    MJD & Flux & $\sigma_{F}$ \\
    &  (Jy) & (Jy)\\
    
    \hline
    
	55845 & 11.7 & 0.96 \\ 
	55921 & 12.2 & 0.97 \\ 
	55921 & 12.0 & 0.96 \\ 
	55922 & 12.1 & 0.97 \\ 
	55922 & 9.84 & 0.81 \\ 
	55922 & 11.9 & 0.96 \\ 
	55923 & 9.81 & 0.81 \\ 
	55923 & 10.9 & 0.89 \\ 
	55924 & 13.0 & 1.05 \\ 
	55924 & 9.90 & 0.82 \\ 
    
    \hline
    \hline
    
    \end{tabular}%
  \label{Table:CWLeo_LCdata_850}%
\end{table}%

\begin{table}
  \centering
  \caption{Sample of $o$ Ceti 850$\micron$ Light Curve Data}
    \begin{tabular}{lll}
    
    \hline
    \hline
    
    MJD & Flux & $\sigma_{F}$ \\
    &  (Jy) & (Jy)\\
    
    \hline
    
55764	&	0.96	&	0.09	\\
55764	&	0.27	&	0.04	\\
55764	&	0.60	&	0.08	\\
55765	&	0.59	&	0.05	\\
55765	&	0.81	&	0.07	\\
55765	&	0.90	&	0.08	\\
55778	&	1.02	&	0.09	\\
55779	&	0.48	&	0.05	\\
55779	&	0.70	&	0.06	\\
55780	&	0.67	&	0.06	\\

    \hline
    \hline
    
    \end{tabular}%
  \label{Table:Mira_LCdata_850}%
\end{table}%

\end{document}